\documentstyle[12pt]{article}
\font\mybb=msbm10 at 12pt
\def\bb#1{\hbox{\mybb#1}}

\def\ZZ {\bb{Z}}
\def\RR {\bb{R}}
\def\CC {\bb{C}}

\font\mybbsubsection=msbm10 at 14pt
\font\mybbabstract=msbm10 at 11pt

\def\bbsubsection#1{\hbox{\mybbsubsection#1}}
\def\bbabstract#1{\hbox{\mybbabstract#1}}

\def\RRsubsection{\bbsubsection{R}}
\def\RRabstract{\bbabstract{R}}

\input{psfig.tex}
\begin{document}
\thispagestyle{empty}
\pagestyle{plain}

\begin{titlepage}
\bigskip
\hskip 4.5in\vbox{\baselineskip12pt
\hbox{INS-Rep.-1191}}\\
\bigskip
\hskip 4.8in\vbox{\baselineskip12pt
\hbox{(revised)}}
\bigskip\bigskip\bigskip\bigskip\\
\centerline{\Large\bf $E_{10}$ Symmetry}
\vskip 5mm
\centerline{\Large\bf in One-dimensional Supergravity} 

\bigskip\bigskip
\bigskip\bigskip

\centerline{Shun'ya Mizoguchi 
\footnote{mizoguch@tanashi.kek.jp}}
\medskip
\centerline{\it Institute of Particle and Nuclear Studies}
\centerline{\it High Energy Accelerator Research Organization, KEK}
\centerline{\it Tanashi, Tokyo 188 Japan}
\bigskip\bigskip

\begin{abstract}
\baselineskip=16pt
We consider dimensional reduction of the eleven-dimensional supergravity 
to less than four dimensions. The three-dimensional $E_{8(+8)}/SO(16)$ 
nonlinear sigma model is derived by direct dimensional reduction from 
eleven dimensions. In two dimensions we explicitly check that the 
Matzner-Misner-type $SL(2,\RRabstract)$ symmetry, together with the 
$E_8$, satisfies the generating relations of $E_9$ under the generalized 
Geroch compatibility (hypersurface-orthogonality) condition. We further 
show that an extra $SL(2,\RRabstract)$ symmetry, which is newly present 
upon reduction to one dimension, extends the symmetry algebra to 
a real form of $E_{10}$. The new $SL(2,\RRabstract)$ acts on certain 
plane wave solutions propagating at the speed of light.  
To show that this $SL(2,\RRabstract)$ cannot be expressed in terms of 
the old $E_9$ but truly enlarges the symmetry, we compactify the final 
two dimensions on a two-torus and confirm that it changes the conformal 
structure of this two-torus. 
\end{abstract} 
\end{titlepage}

\baselineskip=18pt
\setcounter{footnote}{0}
\def\beq{\begin{equation}}
\def\eeq{\end{equation}}
\def\beqa{\begin{eqnarray}}
\def\eeqa{\end{eqnarray}}
\def\n{\nonumber\\}
\def\ld{\lambda}
\def\sm{\sigma}

\def\mb{\overline{\mu}}
\def\nb{\overline{\nu}}
\def\ab{\overline{\alpha}}
\def\beb{\overline{\beta}}

\def\ep{e_{\dot{+}}^+}
\def\emi{e_{\dot{-}}^-}

\def\pld{\dot{+}}
\def\mnd{\dot{-}}

\def\theequation{\arabic{section}.\arabic{equation}}

\section{Introduction}
It is now widely believed that there exists a fundamental 
eleven-dimensional quantum theory which incorporates all 
five superstring theories. Various duality symmetries of 
compactified superstrings \cite{S,T,HT} are discrete subgroups of 
``hidden symmetries'' of their effective supergravity theories. An 
interesting picture is that there is some huge discrete symmetry 
of the most symmetric vacuum of M theory, and other dualities 
arise according to the variety of symmetry breaking. This 
fundamental symmetry would then include all known duality groups 
as subgroups.

The eleven-dimensional supergravity \cite{CJS} possesses 
(continuous) $E_{7(+7)}$ global (with $SU(8)$ local) symmetry \cite{CJ}
when it is compactified to four dimensions on a seven torus $T^7$. 
Its discrete subgroup $E_{7(+7)}$(\ZZ) is known as U-duality
of the compactified typeII superstring \cite{HT}. 
Below four dimensions the eleven-dimensional supergravity 
exhibits $E_8$ and $E_9$ in three and two dimensions
\cite{JuliaCambridge,JuliaErice,MarcusSchwarz,BM,NicolaiE9} with     
the compactification spaces being $T^8$ and $T^9$, respectively.
One is then curious about what happens if one further goes to 
one dimension. In 1982 Julia conjectured that the symmetry group 
will be enlarged to $E_{10}$ \cite{JuliaChicago}, 
whose Lie algebra belongs to a certain class of 
Kac-Moody algebra, ``hyperbolic Kac-Moody algebra''. 
This paper is an attempt to answer to this question.

\begin{figure} 
\caption{Dynkin diagram of $E_{10}$.}
\mbox{\phantom{space}}\\
\centerline{
\mbox{\psfig{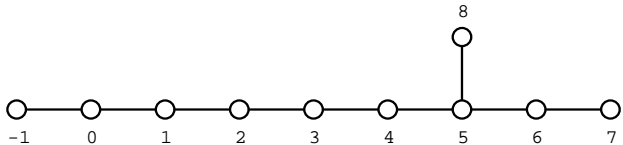}}
}
\label{E10Dynkin}
\end{figure}

The hyperbolic Kac-Moody algebra $E_{10}$ is 
defined by the Cartan matrix 
\beq
K_{ij}=\left[
\begin{array}{ccccccccccc}
2&-1&&&&&&&&\\
\!\!-1\!\!&2&\!\!-1\!\!&&&&&&&\\
&\!\!-1\!\!&2&\!\!-1\!\!&&&&&&\\
&&\!\!-1\!\!&2&\!\!-1\!\!&&&&&\\
&&&\!\!-1\!\!&2&\!\!-1\!\!&&&&\\
&&&&\!\!-1\!\!&2&\!\!-1\!\!&&&\\
&&&&&\!\!-1\!\!&2&\!\!-1\!\!&&\!\!-1\!\!\\
&&&&&&\!\!-1\!\!&2&\!\!-1\!\!&\\
&&&&&&&\!\!-1\!\!&2&\\
&&&&&&\!\!-1\!\!&&&2
\end{array}
\right],
\eeq
whose Dynkin diagram is shown in Fig{\ref{E10Dynkin}}. 
In general one can define a Kac-Moody algebra associated with 
any $N\times N$ Cartan matrix $K_{ij}$ that satisfies 
(i) $K_{ii}=2$, (ii) $K_{ij}$ $(i\neq j)$ is non-positive integer and 
(iii) if $K_{ij}=0$ then $K_{ji}=0$. 
Given such a matrix $K_{ij}$, 
the algebra is defined as arbitrary number of multiple commutators 
with the relations 
\beq
[h_i,~e_j]=K_{ij}e_j,~~[h_i,~f_j]=-K_{ij}f_j,~~
[e_i,~f_j]=\delta_{ij}h_j,~~[h_i,~h_j]=0
\label{generating_relations}
\eeq
$(i,j=1,\ldots,N)$ and the Serre relations \cite{Kac}
\beq
(\mbox{ad}e_i)^{1-K_{ij}}e_j=0,~~(\mbox{ad}f_i)^{1-K_{ij}}f_j=0.
\label{Serre_relation}
\eeq 
The set of generators ${e_i,f_i,h_i}$ $(i=1,\ldots,N)$ 
are called the Chevalley generators.
A Kac-Moody algebra is said hyperbolic if a removal of any vertices from 
its Dynkin diagram leaves diagrams of either finite-dimensional 
simple Lie algebras, affine Kac-Moody algebras or their direct sum. 
A well-known intriguing feature of hyperbolic Kac-Moody algebras, 
in particular in relation to supergravity and string theory, is that 
hyperbolic Kac-Moody algebras can exist only up to $\mbox{rank}=10$. 
$E_{10}$ is one of three hyperbolic Kac-Moody algebras 
with the possible highest rank (See \cite{E10} for a review.). 
One is tempted to suspect \cite{JuliaChicago} that this restriction 
might be linked in some way to the fact that the allowed highest 
dimensions for supergravity is eleven.  

%%%%%%%%%%%%%%%%%%%%%%%%%%%%%%%%%%%%%%%%%%%%%%%%%%%%%%%%%%%%%%%%%
%%%%%%%%%%%%%%%%%%%%%%%%%%%%%%%%%%%%%%%%%%%%%%%%%%%%%%%%%%%%%%%%%

Toward the proof of the $E_{10}$ symmetry, an important progress 
was brought about in 1991 by Nicolai, who pointed out a special 
feature of dimensional reduction to one dimension \cite{NicolaiPL}.   
He showed in $D=4$, $N=1$ supergravity that one must take a null 
Killing vector at the stage of the reduction from two to one 
dimension if one wants to keep the duality relations non-trivial. 
After this novel type of reduction 
\footnote{See \cite{JN} for other aspects of null reduction.}, 
he wrote out for the first time the set of the action of 
the Chevalley generators on the fields 
of the $SL(2,\RR)$ symmetry which newly emerged in one dimension, 
confirming that this together with the Lie algebra of the Geroch 
group generates the hyperbolic ``$SL(2,\RR)$ double-hat'' algebra.

What was subtle in his argument was the independence of the 
new $SL(2,\RR)$. Suppose that one starts with a conformal-gauge 
two-metric (i.e. a zweibein proportional to the identity) as in 
ref.\cite{NicolaiPL} by invoking the degrees of freedom of 
diffeomorphisms. Introducing a constant Killing vector then leads it 
to a flat metric, which is essentially unique on a plane
\footnote{In other words, ``conformal gauge'' is not just a gauge 
choice any more once if the Killing vector is fixed.}. Hence the new 
$SL(2,\RR)$ can do nothing more than the old Geroch group can on the 
two-metric because a global scale transformation can be realized by 
the central charge of the Geroch group 
\footnote{One cannot exclude 
the possibility that it might act non-trivially on other fields than 
the two-metric, so that it might really enlarge the symmetry. 
We will seek another possibility in the following, however.}.  
For this reason, although the $E_9$ symmetry has been known for some 
time, the straightforward-looking generalization to $E_{10}$  
has never been accomplished.

In this paper, to achieve the non-trivial realization of ``Nicolai's 
$SL(2,\RR)$'', we parameterize the two-metric by Lorentzian analogue 
of Beltrami differentials and wrap the final two dimensions on a 
two-torus. In fact, even when a non-conformally-flat metric has been 
chosen, it necessarily becomes flat again once a constant null Killing 
vector is introduced. 
The escape in our case is that the new $SL(2,\RR)$ can act 
on the conformal structure \footnote{The idea of the possibility was 
suggested by H.Nicolai (private communication).} of the two-torus. 
One may view these solutions as certain plane wave solutions 
propagating at the speed of light. We will show that the new 
$SL(2,\RR)$ does cause a change of a ``Beltrami differential'' of 
the Lorentzian metric that cannot be absorbed by any diffeomorphisms.

Another issue in establishing the $E_{10}$ symmetry is how 
the fields of the eleven-dimensional supergravity parameterize 
the symmetric space $E_{8(+8)}/SO(16)$ after the reduction 
to three dimensions. Although the three-dimensional locally 
supersymmetric $E_{8(+8)}/SO(16)$ nonlinear sigma model is 
well-known \cite{MarcusSchwarz}, the direct construction from 
eleven dimensions by dimensional reduction seems to have never 
appeared in print. Of course one could also start from 
$E_{7(+7)}/SU(8)$ in four dimensions \cite{JuliaErice}, 
but the way $E_{7(+7)}/SU(8)$ is embedded into 
$E_{8(+8)}/SO(16)$ is a bit complicated. In this paper 
we obtain the $E_{8(+8)}/SO(16)$ nonlinear sigma model 
by direct dimensional reduction from eleven to three dimensions.   
We use Freudenthal's classical realization of $E_8$ 
\cite{Sur_le_E8,JuliaCambridge}. It turns out that this realization 
clearly reflects the (bosonic) field content in three dimensions 
after duality transformations, showing how the eleven-dimensional 
fields fit into the $E_{8(+8)}/SO(16)$ scalar manifold.

Our strategy for the construction of $E_{10}$ is as follows. 
We proceed step by step. We first read off the set of 
transformations that correspond to the Chevalley generators of 
$E_{8(+8)}$ in the $E_{8(+8)}/SO(16)$ sigma model. Next we reduce 
the dimension to two, write out the Chevalley generators of 
the extra $SL(2,\RR)$ in two dimensions, and confirm that this 
$SL(2,\RR)$ has correct commutation relations with the $E_8$ so that  
the symmetry is extended to $E_9$. Finally we reduce one more 
dimension, find the new $SL(2,\RR)$ transformation, and 
check that it successfully enlarges the symmetry algebra to $E_{10}$
\footnote{Note that if the relations (\ref{generating_relations}) 
and (\ref{Serre_relation}) are regarded as those of a {\em real} 
Lie algebra, with $K_{ij}$ defined by (\ref{E8Cartan}), 
they define a real form $E_{8(+8)}$ of the complex 
Lie algebra $E_8$. (The number $+8$ in parenthesis denotes 
the difference between the numbers of positive and negative 
generators in the invariant bilinear form.) For any simple Lie 
algebra $X_N$, the real Lie algebra defined in this way 
is always $X_{N(+N)}$, since positive and negative roots can be 
paired either symmetrically or anti-symmetrically, giving 
the same number of generators with opposite signs. This 
leaves $N$ excess coming from the Cartan subalgebra.  
In that sense the symmetries should be called $E_{9(+9)}$ and 
$E_{10(+10)}$, respectively, although both are infinite-dimensional.
They are called `normal' (or `split') real forms.}.

This paper is organized as follows. In sect.2 we derive the 
$E_{8(+8)}/SO(16)$ nonlinear sigma model by dimensional reduction 
from eleven dimensions and read off the transformations 
corresponding to the Chevalley generators. 
In sect.3 we deal with the reduced two-dimensional model and 
check that the symmetry is enlarged to $E_9$. In sect.4 we 
reduce the dimension to one and show that the symmetry is $E_{10}$. 
The non-triviality of Nicolai's $SL(2,\RR)$ is also proved. 
Finally we give conclusions in sect.5.

\section{$E_{8}$ in three dimensions}
\setcounter{equation}{0}
\subsection{$E_{8}/SO(16)$ nonlinear sigma model from 
the eleven- dimensional supergravity}
We start with dimensional reduction of the eleven-dimensional 
supergravity \cite{CJS} to three dimensions. 
The bosonic Lagrangian with all fermionic fields set to zero is 
\beqa
{\cal L}&=&
E^{(11)}\left[R^{(11)}-\frac{1}{12}F_{\hat{\mu}\hat{\nu}\hat{\rho}\hat{\sm}}
F^{\hat{\mu}\hat{\nu}\hat{\rho}\hat{\sm}}\right]\n
&&+\frac{8}{12^4}
\epsilon^{\hat{\mu}_1\hat{\mu}_2\hat{\mu}_3\hat{\mu}_4
\hat{\mu}_5\hat{\mu}_6\hat{\mu}_7\hat{\mu}_8
\hat{\mu}_9\hat{\mu}_{10}\hat{\mu}_{11}
}
F_{\hat{\mu}_1\hat{\mu}_2\hat{\mu}_3\hat{\mu}_4}
F_{\hat{\mu}_5\hat{\mu}_6\hat{\mu}_7\hat{\mu}_8}
A_{\hat{\mu}_9\hat{\mu}_{10}\hat{\mu}_{11}}.
\eeqa
$\hat{\mu}$, $\hat{\nu}$, $\ldots$ are the eleven-dimensional 
spacetime indices. Decomposing $x^{\hat{\mu}}$ 
into uncompactified coordinates $x^{\mu}=t,x,y$ and 
compactified coordinates $x^i$, $i=1,\ldots,8$   
and dropping the $x^i$-dependence, we get the three-dimensional 
Lagrangian  
\beqa
{\cal L}&=&
E^{(3)}\left[R^{(3)}-G^{(3)\mu\nu}\partial_{\mu}\ln e\partial_{\nu}\ln e
+\frac{1}{4}G^{(3)\mu\nu}\partial_{\mu}g^{ij}\partial_{\nu}g_{ij}
\right.\n
&&-\frac{1}{4}G^{(3)\mu\rho}G^{(3)\nu\sm}e^2g_{ij}B^i_{\mu\nu}B^j_{\rho\sm}
-\frac{1}{3}G^{(3)\mu\nu}g^{il}g^{jm}g^{kn}\partial_{\mu}A_{ijk}
\partial_{\nu}A_{lmn}\n
&&\left.-\frac{1}{2}G^{(3)\mu\rho}G^{(3)\nu\sm}e^2g^{il}g^{jm}
(F'_{\mu\nu ij}+B^k_{\mu\nu}A_{ijk})(F'_{\rho\sm lm}+B^n_{\rho\sm}A_{lmn})
\right]\n
&&+\frac{1}{36}
\epsilon^{\mu\nu\rho}\epsilon^{ijklmnpq}
(F'_{\mu\nu ij}+\frac{2}{3}B^r_{\mu\nu}A_{ijr})\partial_{\rho}A_{klm}A_{npq},
\label{3dLag}
\eeqa
where
\beq
F'_{\mu\nu ij}=2\partial_{[\mu}A'_{\nu]ij},~~~
A'_{\mu ij}=A_{\mu ij}-B_{\mu}^k A_{ijk},
\eeq
\beq
B^i_{\mu\nu}=2\partial_{[\mu}B_{\nu]}^{i}.
\eeq
$B_{\mu}^{~i}$ is the Kaluza-Klein vector fields associated with the 
decomposition of the elfbein
\beq
E_{~~~\hat{\mu}}^{(11)\hat{\alpha}}
=\left[
\begin{array}{cc}
e^{-1}E_{\mu}^{(3)\alpha}&B_{\mu}^{~i}e_i^{~a}\\
0&e_i^{~a}
\end{array}
\right].\label{3+8}
\eeq
The local Lorentz flat metric is 
\beq
\eta_{\hat{\alpha}\hat{\beta}}=
\left[
\begin{array}{cc}
\eta^{(3)}_{\alpha\beta}&\\
&\delta_{ab}
\end{array}
\right],
\label{eta}
\eeq
where the eleven-dimensional Lorentz indices $\hat{\alpha}$ are 
also decomposed into $\alpha=0,\ldots,3$ and $a=1,\ldots,8$ similarly. 
$\eta^{(3)}_{\alpha\beta}$ has signature $(-++)$.
$A'_{\mu ij}$ is an invariant combination under Kaluza-Klein 
gauge transformations. 

The degrees of freedom of vector fields in three dimensions can be  
traded by those of scalar fields by duality transformations. 
To see this, a usual trick is to introduce the 
following Lagrange multiplier terms 
\beq
{\cal L}_{\mbox{\scriptsize Lag.mul.}}=
\varphi^{ij}\epsilon^{\mu\nu\rho}\partial_{\mu}F'_{\nu\rho ij}
+\frac{1}{2}\psi_i\epsilon^{\mu\nu\rho}\partial_{\mu}B^i_{\nu\rho}.
\label{LLag1}
\eeq
Up to complete squares of $F'_{\mu\nu ij}$ and $B^i_{\mu\nu}$, 
we find 
\beqa
&&{\cal L}+{\cal L}_{\mbox{\scriptsize Lag.mul.}}\n
&=&
E^{(3)}\left[R^{(3)}
+G^{(3)\mu\nu}\left(
\frac{1}{4}\partial_{\mu}g^{ij}\partial_{\nu}g_{ij}
-\partial_{\mu}\ln e\partial_{\nu}\ln e
-\frac{1}{3}g^{il}g^{jm}g^{kn}\partial_{\mu}A_{ijk}
\partial_{\nu}A_{lmn}\right.\right.\n
&&
~~~~~~~-e^2g_{ik}g_{jl}(\partial_{\mu}\varphi^{ij}
-\frac{1}{36}\epsilon^{ijp_1q_1r_1s_1t_1u_1}\partial_{\mu}
A_{p_1q_1r_1}A_{s_1t_1u_1})\n
&&~~~~~~~~~~~~~~~~~~~\cdot(\partial_{\nu}\varphi^{kl}
-\frac{1}{36}\epsilon^{klp_2q_2r_2s_2t_2u_2}\partial_{\nu}
A_{p_2q_2r_2}A_{s_2t_2u_2})\n
&&
~~~~~~~-\frac{1}{2}e^{-2}g^{ij}
(\partial_{\mu}\psi_i+\partial_{\mu}A_{ik_1l_1}\varphi^{k_1l_1}
-A_{ik_1l_1}\partial_{\mu}\varphi^{k_1l_1}\n&&~~~~~~~~~~~~~~~~~~~
+\frac{1}{54}\epsilon^{m_1n_1p_1q_1r_1s_1t_1u_1}
A_{im_1n_1}\partial_{\mu}A_{p_1q_1r_1}A_{s_1t_1u_1})\n
&&~~~~~~~~~~~~~~~~~~~
\cdot(\partial_{\nu}\psi_j+\partial_{\nu}A_{jk_2l_2}\varphi^{k_2l_2}
-A_{ik_2l_2}\partial_{\nu}\varphi^{k_2l_2}\n&&~~~~~~~~~~~~~~~~~~~
\left.\left.+\frac{1}{54}\epsilon^{m_2n_2p_2q_2r_2s_2t_2u_2}
A_{im_2n_2}\partial_{\nu}A_{p_2q_2r_2}A_{s_2t_2u_2})\right)\right],
\label{reducedL}
\eeqa
where the degrees of freedom of $28+8$ three-dimensional vectors 
$A'_{\mu ij}$ and $B_{\mu}^i$ are replaced by those of the scalars 
$\varphi^{ij}$ and $\psi_i$. Duality relations are 
\beq
F^{(3)\mu\nu ij}=-E^{(3)-1}e^{-2}\epsilon^{\mu\nu\rho}
(\partial_{\rho}\varphi^{ij}-\frac{1}{36}\epsilon^{ijklmnpq}
\partial_{\rho}A_{klm}~A_{npq}) \label{duality1}
\eeq
with $F^{(3)}_{\mu\nu ij}=F'_{\mu\nu ij}+B_{\mu\nu}^kA_{ijk}$,
and
\beqa
B_k^{\mu\nu}&=&-E^{(3)-1}e^{-2}\epsilon^{\mu\nu\rho}
(\partial_{\rho}\psi_k-A_{kij}\partial_{\rho}\varphi^{ij}
+\varphi^{ij}\partial_{\rho}A_{kij}\n
&&~~~~~~~~~~~~~~~~~~~~~~
+\frac{1}{54}\epsilon^{ijlmnpqr}A_{ijk}\partial_{\rho}A_{lmn}~A_{pqr}).
\label{duality2}
\eeqa
The derivation of (\ref{reducedL}) is standard. 

We will now show how this nonlinear sigma model fits into the 
symmetric space $E_{8(+8)}/SO(16)$. We first observe
that if all the fields originating from 
$A_{\hat{\mu}\hat{\nu}\hat{\rho}}$ are turned off, 
the remaining scalars describe 
$SL(9,\RR)/SO(9)$. This can be seen as follows. 
The sigma-model part of 
the Lagrangian (\ref{reducedL}) is then simply given by 
 
\beq
E^{(3)}G^{(3)\mu\nu}\left(
\frac{1}{4}\partial_{\mu}g^{ij}\partial_{\nu}g_{ij}
-\partial_{\mu}\ln e\partial_{\nu}\ln e
-\frac{1}{2}e^2g^{ij}\partial_{\mu}\psi_i\partial_{\nu}\psi_j
\right).\label{sl9Rsigmamodel}
\eeq
We parameterize the coset $SL(9,\RR)/SO(9)$
in terms of a nine-by-nine symmetric matrix $M\equiv VV^T$ with,
taking $e_i^{~a}$ to be upper-triangular, 
\beq
V_{i'}^{~~a'}=\left[\begin{array}{cc}
e_i^{~a}& -e^{-1}\psi_i\\0&e^{-1}
\end{array}\right].
\eeq
Here we introduced the extended curved and flat indices $i'=i,9$ 
and $a'=a,9$ in a similar way to \cite{CJ} where the 
$SL(7,\RR)$ is enlarged to the $SL(8,\RR)$. 
The form of $VV^T$ ensures that the orthogonal group 
factor is gauged out. One may then easily see that 
(\ref{sl9Rsigmamodel}) can be written 
\beq
\frac{1}{4}E^{(3)}G^{(3)\mu\nu}\mbox{Tr}\partial_{\mu}M^{-1}\partial_{\nu}M.
\eeq                
This is an example of a general rule that $d$-dimensional pure gravity 
exhibits $SL(d-2,\RR)/SO(d-2)$ symmetry when it is reduced to 
three dimensions \cite{JuliaCambridge}. Another (the oldest) example is 
``Ehlers' $SL(2,\RR)$'' \cite{Ehlers,BM} known for a long time 
in general relativity.

As a complex Lie algebra, $SL(9,\CC)$ can be extended to $E_8$ 
by adding two third-rank antisymmetric tensors 
\cite{Sur_le_E8}. Let $V$ be the 
space of third-rank antisymmetric tensor representation of $SL(9,\CC)$,
and $V^*$ be the dual space of $V$, $E_8$ is decomposed as 
$SL$(9,{\bf C})$\oplus V \oplus V^*$. Their dimensions are 
\mbox{\bf $248=80\oplus 84 \oplus \overline{84}$}. $E_8$ can be defined 
by the following commutation relations (Lie brackets):
\beqa
&&[X, ~~Y]_{i'}^{~j'}=X_{i'}^{~k'}Y_{k'}^{~j'}-X_{k'}^{~j'}Y_{i'}^{~k'},\n
&&[X,~~v]_{i'j'k'}
=X_{i'}^{~l'}v_{l'j'k'}+X_{j'}^{~l'}v_{i'l'k'}+X_{k'}^{~l'}v_{i'j'l'},\n
&&[X,~~v^*]^{i'j'k'}
=-(X_{l'}^{~i'}v^{*l'j'k'}+X_{l'}^{~j'}v^{*i'l'k'}+X_{l'}^{~k'}v^{*i'j'l'}),\n
&&[v,~~w]^{i'j'k'}
=\frac{1}{36\sqrt{3}}\epsilon^{i'j'k'p'q'r'l'm'n'}v_{p'q'r'}w_{l'm'n'},\n
&&[v^*,~w^*]_{i'j'k'}
=\frac{1}{36\sqrt{3}}\epsilon_{i'j'k'p'q'r'l'm'n'}v^{*p'q'r'}w^{*l'm'n'},\n
&&[v,~~w^*]_{i'}^{~j'}
=-\frac{1}{6}(v_{i'k'l'}w^{*j'k'l'}
                         -\frac{1}{9}\delta_{i'}^{j'}v_{k'l'm'}w^{*k'l'm'}),\n
\eeqa
$X_{i'}^{~j'},Y_{i'}^{~j'}\in$ $SL(9,\CC)$, 
$v_{i'j'k'},w_{i'j'k'}\in V$ and  $v^{*i'j'k'},w^{*i'j'k'}\in V^*$.
If these relations are regarded as those of a real Lie algebra, 
then they define  $E_{8(+8)}$.
We write down the generators in adjoint representation  
 of $E_{8(+8)}$ in the three-by-three block form
\beqa
\mbox{ad}X&=&\left[
\begin{array}{ccc}
X_{i'}^{~l'}\delta_{m'}^{j'}-X_{m'}^{j'}\delta_{i'}^{~l'}&0&0\\
&&\\
0&~~~~~~~~X_{i'j'k'}^{l'm'n'}&0\\
&&\\
0&0&~~~~~~~~-X_{l'm'n'}^{i'j'k'}~~~
\end{array}\right],\n
\mbox{ad}v~&=&\left[
\begin{array}{ccc}
0&0&\mbox{\scriptsize 
$-\frac{1}{6}(v_{i'[m'n'}\delta_{l']}^{j'}
  -\frac{1}{9}v_{l'm'n'}\delta_{i'}^{j'})$}\\
&&\\
\mbox{\scriptsize $-3(v_{m'[j'k'}\delta_{i']}^{l'}
-\frac{1}{9}v_{i'j'k'}\delta_{m'}^{l'})
$}&0&0\\
&&\\
0&\mbox{\scriptsize 
$\frac{1}{36\sqrt{3}}\epsilon^{i'j'k'p'q'r'l'm'n'}v_{p'q'r'}$}&0
\end{array}\right],\n
\mbox{ad}v^*&=&\left[
\begin{array}{ccc}
0&\mbox{\scriptsize 
$\frac{1}{6}(v^{*j'[m'n'}\delta^{l']}_{i'}
  -\frac{1}{9}v^{*l'm'n'}\delta_{i'}^{j'})$}&0\\
&&\\
0&0&\mbox{\scriptsize 
$\frac{1}{36\sqrt{3}}\epsilon_{i'j'k'p'q'r'l'm'n'}v^{*p'q'r'}$}\\
&&\\
\mbox{\scriptsize $3(v^{*l'[j'k'}\delta^{i']}_{m'}
-\frac{1}{9}v^{*i'j'k'}\delta^{l'}_{m'})$}&0&0
\end{array}\right],\n
\eeqa
with
$X_{i'j'k'}^{l'm'n'}
\equiv 
X_{i'}^{~l'}\delta_{j'}^{m'}\delta_{k'}^{n'}
+X_{j'}^{~l'}\delta_{k'}^{m'}\delta_{i'}^{n'}
+X_{k'}^{~l'}\delta_{i'}^{m'}\delta_{j'}^{n'}$. 
They map tensors $(Y_{l'}^{m'}$, $w_{l'm'n'}$, $w^{*l'm'n'})$
to $(Y'_{i'}{}^{j'}$, $w'_{i'j'k'}$, $w'{}^{*i'j'k'})$.

In view of the structure of the algebra, 
$56+28=84$ $A_{ijk}$ and $\varphi^{ij}$ should fit somehow into 
$84+84$ $v_{i'j'k'}$ and $v^{*i'j'k'}$. To find the correct assignment, 
a useful hint is that $\mbox{ad}(v+v^*)$ must be a nilpotent matrix so that 
$\exp(\mbox{ad}(v+v^*))$ becomes a polynomial. The most natural candidate 
is the case that all $A_{ijk}$ and $\varphi^{ij}$ correspond only to 
the elements associated with negative (or positive) roots. 
In this way a half of $84+84$ elements 
are selected. Moreover, it turns out that   
$v_{ijk}$ and $v^{*ij9}$ $(i,j,k=1,\ldots,8)$ 
can be taken as the elements 
corresponding to negative roots. In view of this nice index structure,
we assume 
\beqa
v_{i'j'k'}&\equiv&A_{i'j'k'}
=-2\sqrt{3}\delta_{[i'}^i\delta_{j'}^j\delta_{k']}^k A_{ijk},\n
v^{*i'j'k'}&\equiv&\varphi^{i'j'k'}
=-2\sqrt{3}\cdot3\delta^{[i'}_i\delta^{j'}_j\delta^{k']}_{9} \varphi^{ij},
\label{normalization}
\eeqa
where $-2\sqrt{3}$ is a normalization for later convenience.
One may then verify that $(\mbox{ad}(v+v^*))^5=0$. Therefore 
$\exp(\mbox{ad}(v+v^*))$ is a fourth-order polynomial of $A_{i'j'k'}$  
and $\varphi^{i'j'k'}$. After some calculation we obtain
\beqa
&&\exp(\mbox{ad}(v+v^*))\n
&=&\left[\begin{array}{lll}
\exp(\mbox{ad}(v+v^*))_{Y,Y}&
\exp(\mbox{ad}(v+v^*))_{Y,w}&
\exp(\mbox{ad}(v+v^*))_{Y,w^*}\\
\exp(\mbox{ad}(v+v^*))_{w,Y}&
\exp(\mbox{ad}(v+v^*))_{w,w}&
\exp(\mbox{ad}(v+v^*))_{w,w^*}\\
\exp(\mbox{ad}(v+v^*))_{w^*,Y}&
\exp(\mbox{ad}(v+v^*))_{w^*,w}&
\exp(\mbox{ad}(v+v^*))_{w^*,w^*}
\end{array}\right],\n
\eeqa
\beqa
&&\exp(\mbox{ad}(v+v^*))_{Y,Y}\n
&=&\delta_{i'}^{l'}\delta_{m'}^{j'}
-\frac{1}{9}\delta_{i'}^{j'}\delta_{m'}^{l'}
-\frac{1}{12}(A_{i'p'q'}\varphi^{l'p'q'}\delta_{m'}^{j'}
+A_{m'p'q'}\varphi^{j'p'q'}\delta_{i'}^{l'}
\n&&+4A_{i'm'p'}\varphi^{j'l'p'}
-\frac{2}{3}A_{m'p'q'}\varphi^{l'p'q'}\delta_{i'}^{j'}
-\frac{2}{3}A_{i'p'q'}\varphi^{j'p'q'}\delta_{m'}^{l'})\n
&&+\frac{1}{432\sqrt{3}}(\epsilon^{j'w'x'p'q'r'l'u'v'}
A_{i'w'x'}A_{p'q'r'}A_{m'u'v'}
+\epsilon_{i'w'x'p'q'r'm'u'v'}
\varphi^{j'w'x'}\varphi^{p'q'r'}\varphi^{l'u'v'})\n
&&-\frac{1}{144}\varphi^{j'p'q'}\varphi^{l'u'v'}
(A_{i'p'q'}A_{m'u'v'}-4A_{i'p'v'}A_{m'u'q'}),\n
&&\\
%%%%%%%
&&\exp(\mbox{ad}(v+v^*))_{Y,w}\n
&=&\frac{1}{6}(\varphi^{j'[m'n'}\delta_{i'}^{l']}
-\frac{1}{9}\delta_{i'}^{j'}\varphi^{l'm'n'})
-\frac{1}{432\sqrt{3}}\epsilon^{j'v'w'p'q'r'l'm'n'}A_{i'v'w'}A_{p'q'r'}\n
&&+\frac{1}{36}\left(\frac{1}{6}A_{i'p'q'}\varphi^{j'p'q'}\varphi^{l'm'n'}
               -A_{i'p'q'}\varphi^{j'p'[l'}\varphi^{m'n']q'}
               +(A_{p'q'r'}
               \varphi^{p'q'l'}
               \varphi^{m'j'r'}
               \delta_{i'}^{n'})^{[l'm'n']}\right)\n
&&+\frac{1}{432\sqrt{3}}\epsilon^{l'm'n'p'q'r's't'u'}
\left(\frac{1}{3}A_{i'q'r'}A_{p'v'w'}
+\frac{2}{3}A_{i'v'p'}A_{q'r'w'}\right)
A_{s't'u'}\varphi^{j'v'w'},\n
&&\\
%%%%%%%
&&\exp(\mbox{ad}(v+v^*))_{Y,w^*}\n
&=&
-\frac{1}{6}(A_{i'[m'n'}\delta^{j'}_{l']}
-\frac{1}{9}\delta_{i'}^{j'}A_{l'm'n'})
+\frac{1}{432\sqrt{3}}\epsilon_{i'v'w'p'q'r'l'm'n'}
                 \varphi^{j'v'w'}\varphi^{p'q'r'}\n
&&-\frac{1}{36}\left(\frac{1}{6}\varphi^{j'p'q'}A_{i'p'q'}A_{l'm'n'}
               -\varphi_{j'p'q'}A_{i'p'[l'}A_{m'n']q'}\right)\n
&&+\frac{1}{432\sqrt{3}}\epsilon^{j'v'w'p'q'r's't'u'}
A_{i'v'w'}A_{p'[m'n'}A_{l']q'r'}A_{s't'u'},
\n
&&\\
%%%%%%%
&&\exp(\mbox{ad}(v+v^*))_{w,Y}\n
&=&
-3(A_{m'[j'k'}\delta_{i']}^{l'}-\frac{1}{9}A_{i'j'k'}\delta_{m'}^{l'})
+\frac{1}{24\sqrt{3}}\epsilon_{i'j'k'p'q'r'm'u'v'}
\varphi^{l'u'v'}\varphi^{p'q'r'}\n
&&+\frac{1}{6}(-\frac{1}{2}\varphi^{l'p'q'}A_{m'p'q'}A_{i'j'k'}
+3\varphi^{l'p'q'}A_{m'p'[i'}A_{j'k']q'})\n
&&-\frac{1}{24\sqrt{3}}\epsilon^{l'v'w'p'q'r's't'u'}
A_{m'v'w'}A_{p'[j'k'}A_{i']q'r'}A_{s't'u'},\n
&&\\
%%%%%%%
&&\exp(\mbox{ad}(v+v^*))_{w,w}\n
&=&
\delta_{[i'}^{l'}\delta_{j'}^{m'}\delta_{k']}^{n'}
+\frac{1}{2}
 \left(\frac{1}{2}(A_{i'p'q'}\varphi^{l'p'q'}
                          \delta_{j'}^{m'}\delta_{k'}^{n'})_{[i'j'k']}
       -(A_{i'j'p'}\varphi^{l'm'p'}\delta_{k'}^{n'})_{[i'j'k'}
       +\frac{1}{9}A_{i'j'k'}\varphi^{l'm'n'}\right)
\n
&&+\frac{1}{72\sqrt{3}}\epsilon^{l'm'n'p'q'r's't'u'}
  A_{p'[j'k'}A_{i']q'r'}A_{s't'u'},\n
&&\\
%%%%%%%
&&\exp(\mbox{ad}(v+v^*))_{w,w^*}\n
&=&
\frac{1}{36\sqrt{3}}\epsilon_{i'j'k'p'q'r'l'm'n'}\varphi^{p'q'r'}
+\frac{1}{4}\left(A_{l'[j'k'}A_{i']m'n'}
-\frac{1}{9}A_{i'j'k'}A_{l'm'n'}
\right),
\n
&&\\
%%%%%%%
&&\exp(\mbox{ad}(v+v^*))_{w^*,Y}\n
&=&
+3(\varphi^{l'[j'k'}\delta_{m'}^{i']}-
\frac{1}{9}\varphi^{i'j'k'}\delta_{m'}^{l'})
-\frac{1}{24\sqrt{3}}\epsilon^{i'j'k'p'q'r'l'u'v'}
A_{m'u'v'}A_{p'q'r'}\n
&&-\frac{1}{6}(-\frac{1}{2}A_{m'p'q'}\varphi^{l'p'q'}\varphi^{i'j'k'}
+3A_{m'p'q'}\varphi^{l'p'[i'}\varphi^{j'k']q'}
-3(A_{p'q'r'}\varphi^{p'q'i'}\varphi^{j'l'r'}\delta_{m'}^{k'})^{[i'j'k']})\n
&&-\frac{1}{24\sqrt{3}}\epsilon^{i'j'k'p'q'r's't'u'}
\left(
\frac{1}{3}A_{m'q'r'}A_{p'v'w'}+\frac{2}{3}A_{m'v'p'}A_{q'r'w'}
\right)
A_{s't'u'}\varphi^{l'v'w'},\n
&&\\
%%%%%%%
&&\exp(\mbox{ad}(v+v^*))_{w^*,w}\n
&=&
\frac{1}{36\sqrt{3}}\epsilon^{i'j'k'p'q'r'l'm'n'}A_{p'q'r'}
+\frac{1}{4}\left(\varphi^{l'[j'k'}\varphi^{i']m'n'}
-\frac{1}{9}\varphi^{i'j'k'}\varphi^{l'm'n'}
\right),
\n
&&\\
%%%%%%%
&&\exp(\mbox{ad}(v+v^*))_{w^*,w^*}\n
&=&
\delta^{[i'}_{l'}\delta^{j'}_{m'}\delta^{k']}_{n'}
+\frac{1}{2}
 \left(\frac{1}{2}(\varphi^{i'p'q'}A_{l'p'q'}
                          \delta^{j'}_{m'}\delta^{k'}_{n'})^{[i'j'k']}
       -(\varphi^{i'j'p'}A_{l'm'p'}\delta^{k'}_{n'})^{[i'j'k']}
       +\frac{1}{9}\varphi^{i'j'k'}A_{l'm'n'}\right)\n
&&-\frac{1}{72\sqrt{3}}\epsilon^{i'j'k'p'q'r's't'u'}
A_{p'[m'n'}A_{l']q'r'}A_{s't'u'}.
\n
\eeqa
Here the bracket means that the indices 
inside are totally anti-symmetrized.
We have employed the equations $A_{i'j'k'}\varphi^{i'j'k'}=0$, 
$A_{i'l'm'}A_{j'k'n'}\varphi^{i'j'k'}=0$, etc. 
We further calculate 
$\exp(-\mbox{ad}(v+v^*))$ $\cdot\partial_{\mu} \exp(\mbox{ad}(v+v^*))$ 
to find
\beqa
&&\exp(-\mbox{ad}(v+v^*))\partial_{\mu} \exp(\mbox{ad}(v+v^*))\n
&&\n
&=&
\left[
\begin{array}{ccc}
Y_{i'}^{~l'}\delta_{m'}^{j'}-Y_{m'}^{j'}\delta_{i'}^{~l'}&
\frac{1}{6}(w^{*j'[m'n'}\delta^{l']}_{i'}
  -\frac{1}{9}w^{*l'm'n'}\delta_{i'}^{j'})&
-\frac{1}{6}(w_{i'[m'n'}\delta_{l']}^{j'} 
  -\frac{1}{9}w_{l'm'n'}\delta_{i'}^{j'})\\
&&\\
-3(w_{m'[j'k'}\delta_{i']}^{l'}
-\frac{1}{9}w_{i'j'k'}\delta_{m'}^{l'})&
Y_{i'j'k'}^{l'm'n'}&
\frac{1}{36\sqrt{3}}\epsilon_{i'j'k'p'q'r'l'm'n'}w^{*p'q'r'}\\
&&\\
~3(w^{*l'[j'k'}\delta^{i']}_{m'}
-\frac{1}{9}w^{*i'j'k'}\delta^{l'}_{m'})&
\frac{1}{36\sqrt{3}}\epsilon^{i'j'k'p'q'r'l'm'n'}w_{p'q'r'}&
-Y_{l'm'n'}^{i'j'k'}
\end{array}\right],\n
\label{adYww*}
\eeqa
with
\beqa
Y_{i'}^{~l'}&=&\frac{1}{12}
\left(A_{i'p'q'}\partial_{\mu}\varphi^{l'p'q'}
-\varphi^{l'p'q'}\partial_{\mu}A_{i'p'q'}\right)\n&&
-\frac{1}{1296\sqrt{3}}
\epsilon^{l'v'w'p'q'r's't'u'}A_{i'v'w'}A_{p'q'r'}\partial_{\mu}A_{s't'u'},\n
w_{i'j'k'}&=&\partial_{\mu}A_{i'j'k'},\n
w^{*i'j'k'}&=&\partial_{\mu}\varphi^{i'j'k'}
-\frac{1}{72\sqrt{3}}
 \epsilon^{i'j'k'p'q'r's't'u'}A_{p'q'r'}\partial_{\mu}A_{s't'u'}.
\eeqa
All terms higher than fourth-order vanish. This is exactly what 
we want because the Lagrangian  
(\ref{reducedL}) contains bilinear of cubic or lower-order terms only. 
Since a triple $[Y_{i'}^{l'},w_{i'j'k'},w^{*i'j'k'}]$ completely 
specifies an element of $E_{8(+8)}$ in adjoint representation, 
we write 
\beq
(\ref{adYww*})\equiv \mbox{ad}[Y_{i'}^{l'},w_{i'j'k'},w^{*i'j'k'}]
\eeq
for simple notation. 

We will now construct the $E_{8(+8)}/SO(16)$ nonlinear 
sigma-model Lagrangian. Let ${\cal V}$ be 
an element of the Lie group of $E_{8(+8)}$ defined by 
\beq
{\cal V}\equiv {\cal V}_-{\cal V}_+
\eeq
with
\beq
{\cal V}_+=\left[
\begin{array}{ccc}
V_{i'}^{~a'}V^{j'}_{~b'}&&\\
&V_{[i'}^{~a'}V_{j'}^{~b'}V_{k']}^{~c'}&\\
&&V^{[i'}_{~a'}V^{j'}_{~b'}V^{k']}_{~c'}
\end{array}
\right]
\eeq
and
\beq
{\cal V}_-=\exp(\mbox{ad}(v+v^*)).
\eeq
The invariant tangent vector field is 
\beq
{\cal V}^{-1}\partial_{\mu}{\cal V}
=\mbox{ad}[Z_{a'}^{~d'},w_{a'b'c'},w^{*a'b'c'}],
\eeq
where 
\beqa
Z_{a'}^{~d'}&=&\left[\begin{array}{cc}
e_a^{~i}\partial_{\mu}e_i^{~d}&
-e^{-1}e_a^{~i}(\partial_{\mu}\psi_i-Y_i^{~9}) \\
0&-e^{-1}\partial_{\mu}e
\end{array}\right],\\
Y_i^{~9}&=&
A_{ipq}\partial_{\mu}\varphi^{pq}
-\varphi^{pq}\partial_{\mu}A_{ipq}
-\frac{1}{54}
\epsilon^{vwpqrstu}A_{ivw}\partial_{\mu}A_{pqr}A_{stu},\n
&&\\
w_{a'b'c'}&=&-2\sqrt{3}\delta_{[a'}^a\delta_{b'}^b\delta_{c']}^c
e_a^{~i}e_b^{~j}e_c^{~k}\partial_{\mu}A_{ijk},\\
w^{*a'b'c'}&=&-6\sqrt{3}\delta^{[a'}_a\delta^{b'}_b\delta^{c']}_9
e^{-1}e^a_{~i}e^b_{~j}(\partial_{\mu}\varphi^{ij}
-\frac{1}{36}\epsilon^{ijpqrstu}\partial_{\mu}A_{pqr}A_{stu}).
\eeqa

To construct a coset nonlinear sigma model, we introduce the 
symmetric space involution $\tau$ such that
for $E_{8(+8)}=\mbox{\bf H}(=SO(16))\oplus \mbox{\bf K}$,
\beq
\tau(\mbox{\bf H})=+\mbox{\bf H},~~~\tau(\mbox{\bf K})=-\mbox{\bf K}.
\eeq
Defining the ``symmetric matrix'' 
\beq
{\cal M}\equiv{\cal V}\cdot\tau({\cal V})^{-1},
\eeq
one finds  
\beq
\mbox{Tr}\partial_{\mu}{\cal M}^{-1}\partial^{\mu}{\cal M}
=-\mbox{Tr}\left({\cal V}^{-1}\partial_{\mu}{\cal V}
-\tau({\cal V}^{-1}\partial_{\mu}{\cal V})\right)^2.
\eeq
Thus the {\bf H} components of ${\cal V}^{-1}\partial_{\mu}{\cal V}$ 
are projected out.  
Since $\tau$ acts on a triple as 
\beq
\tau(\mbox{ad}[Z_{a'}^{~d'},w_{a'b'c'},w^{*a'b'c'}])
=\mbox{ad}[-Z_{d'}^{~a'},w^*_{a'b'c'},w^{a'b'c'}],
\eeq
we obtain
\beqa
&&\mbox{Tr}\partial_{\mu}{\cal M}^{-1}\partial^{\mu}{\cal M}\n
&=&-\mbox{Tr}(\mbox{ad}
[(Z+Z^T)_{a'}^{~d'},(w-w^*)_{a'b'c'},(w^*-w)^{a'b'c'}]
)^2\n
&=&-60\left((Z+Z^T)_{a'}^{~b'}(Z+Z^T)_{b'}^{~a'}
+\frac{1}{9}(w-w^*)_{a'b'c'}(w-w^*)^{a'b'c'}
\right)\n
&=&240G^{(3)\mu\nu}\left(
\frac{1}{4}\partial_{\mu}g^{ij}\partial_{\nu}g_{ij}
-\partial_{\mu}\ln e\partial_{\nu}\ln e
-\frac{1}{3}g^{il}g^{jm}g^{kn}\partial_{\mu}A_{ijk}
\partial_{\nu}A_{lmn}\right.\n
&&
~~~~~~~~~~~~~~-e^{-2}g_{ik}g_{jl}(\partial_{\mu}\varphi^{ij}
-\frac{1}{36}\epsilon^{ijp_1q_1r_1s_1t_1u_1}\partial_{\mu}
A_{p_1q_1r_1}A_{s_1t_1u_1})\n
&&~~~~~~~~~~~~~~~~~~~~~~~~~~\cdot(\partial_{\nu}\varphi^{kl}
-\frac{1}{36}\epsilon^{klp_2q_2r_2s_2t_2u_2}\partial_{\nu}
A_{p_2q_2r_2}A_{s_2t_2u_2})\n
&&
~~~~~~~~~~~~~~-\frac{1}{2}e^{-2}g^{ij}
(\partial_{\mu}\psi_i+\partial_{\mu}A_{ik_1l_1}\varphi^{k_1l_1}
-A_{ik_1l_1}\partial_{\mu}\varphi^{k_1l_1}\n&&~~~~~~~~~~~~~~~~~~~
+\frac{1}{54}\epsilon^{m_1n_1p_1q_1r_1s_1t_1u_1}
A_{im_1n_1}\partial_{\mu}A_{p_1q_1r_1}A_{s_1t_1u_1})\n
&&~~~~~~~~~~~~~~~~~~~~~~~~~~
\cdot(\partial_{\nu}\psi_j+
\partial_{\nu}A_{jk_2l_2}\varphi^{k_2l_2}
-A_{ik_2l_2}\partial_{\nu}\varphi^{k_2l_2}\n&&~~~~~~~~~~~~~~~~~~~
\left.+\frac{1}{54}\epsilon^{m_2n_2p_2q_2r_2s_2t_2u_2}
A_{im_2n_2}\partial_{\nu}A_{p_2q_2r_2}A_{s_2t_2u_2})\right).
\eeqa
This expression is identical (except for the factor 240) 
to the sigma-model part of (\ref{reducedL}). 
This completes the direct derivation of the $E_{8(+8)}/SO(16)$ 
nonlinear sigma model by dimensional reduction from eleven dimensions.

\subsection{Chevalley generators of $E_8$}
For later convenience we write out the transformation formulas 
of the scalar fields corresponding to the Chevalley generators 
of $E_8$, which satisfy the relations 
(\ref{generating_relations})(\ref{Serre_relation}) 
with the Cartan matrix 
\beq
K_{ij}=\left[
\begin{array}{cccccccc}
2&\!\!-1\!\!&&&&&&\\
\!\!-1\!\!&2&\!\!-1\!\!&&&&&\\
&\!\!-1\!\!&2&\!\!-1\!\!&&&&\\
&&\!\!-1\!\!&2&\!\!-1\!\!&&&\\
&&&\!\!-1\!\!&2&\!\!-1\!\!&&\!\!-1\!\!\\
&&&&\!\!-1\!\!&2&\!\!-1\!\!&\\
&&&&&\!\!-1\!\!&2&\\
&&&&\!\!-1\!\!&&&2
\end{array}
\right]. \label{E8Cartan}
\eeq

\begin{figure} 
\caption{Dynkin diagram of $E_8$.}
\mbox{\phantom{space}}\\
\centerline{
\mbox{\psfig{figure=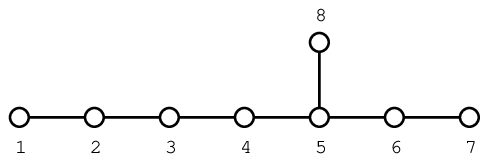,width=6.25cm}}
}
\label{E8Dynkin}
\end{figure}

Labeling the vertices as in Fig.\ref{E8Dynkin}, 
we denote the $k$th $SL(2,\RR)$ subalgebra by $SL(2,\RR)_n$ 
$(n=1,\ldots,8)$. They generate $E_{8(+8)}$ as already remarked.  
Then $SL(2,\RR)_k$ $(k=1,\ldots,7)$ act on ${\cal V}$ as  
(before compensating $SO(16)$ gauge transformations)
\footnote{In fact it is $-\delta_{x_n}$ that satisfy the algebra
of $x_n$ $(x=e,f,h)$ because $[\delta_{x},~\delta_{x'}]=\delta_{[x',x]}$.} 
\beqa
\delta_{e_k}{\cal V}&=&\mbox{ad}[E_{(k)i'}^{~~~~l'},0,0]{\cal V},\n
\delta_{f_k}{\cal V}&=&\mbox{ad}[F_{(k)i'}^{~~~~l'},0,0]{\cal V},\n
\delta_{h_k}{\cal V}&=&\mbox{ad}[H_{(k)i'}^{~~~~l'},0,0]{\cal V},
\eeqa
with 
\beq
E_{(k)i'}^{~~~~l'}=\delta_{i'}^k\delta_{k+1}^{l'},~~
F_{(k)i'}^{~~~~l'}=\delta_{i'}^{k+1}\delta_k^{l'},~~
H_{(k)i'}^{~~~~l'}
=\delta_{i'}^k\delta_k^{l'}-\delta_{i'}^{k+1}\delta_{k+1}^{l'},
\eeq
whereas for $SL(2,\RR)_8$, 
\beqa
\delta_{e_8}{\cal V}&=&\mbox{ad}[0,6\sqrt{3}\delta^{6}_{[i'}
                    \delta^{7}_{j'}
                    \delta^{8}_{k']},0]{\cal V},\n
\delta_{f_8}{\cal V}&=&
     \mbox{ad}[0,0,-6\sqrt{3}\delta_{6}^{[i'}
                    \delta_{7}^{j'}
                    \delta_{8}^{k']}]{\cal V},\n
\delta_{h_8}{\cal V}&=&\mbox{ad}[\tilde{H}_{(8)i'}
             ^{~~~~l'},0,0]{\cal V},
\eeqa
where
\beqa
\tilde{H}_{(8)}
&\equiv& -\frac{1}{3}(H_{(1)}+2H_{(2)}+3H_{(3)}+4H_{(4)}+5H_{(5)}+3H_{(6)}
\n&&~~~~~+H_{(7)}-H_{(8)}),
\eeqa
\beq
H_{(8)i'}^{~~~~l'}\equiv
\delta_{i'}^{8}\delta_{8}^{l'}
-\delta_{i'}^{9}\delta_{9}^{l'}.
\eeq

%%%%%%%%%%%%%%%%%%%%%%%%%%%%%%%%%%%%%%%%%%%%%%%%%%%%%%%%%%%%%%%%%
%%%%%%%%%%%%%%%%%%%%%%%%%%%%%%%%%%%%%%%%%%%%%%%%%%%%%%%%%%%%%%%%%

Using the equation 
\beq
{\cal V}_-^{-1}\delta{\cal V}_-+\delta{\cal V}_+{\cal V}_+^{-1}
={\cal V}_-^{-1}X{\cal V}_-
\eeq
for $\delta{\cal V}=X{\cal V}$, we may express these formulas as those for
$A_{i'j'k'}$,$\varphi^{i'j'k'}$ and $V_{i'}^{a'}$. The result is as 
follows:  
\beqa
&&\delta_{e_k}A_{i'j'k'}=3A_{k+1[j'k'}\delta_{i']}^k,\n
&&\delta_{e_k}\varphi^{i'j'k'}=-3\varphi^{k[j'k'}\delta^{i']}_{k+1},\\
&&\delta_{e_k}V_{i'}^{a'}=\delta_{i'}^{k}\delta_{k+1}^{l'}V_{l'}^{a'},\n
&&\n
&&\delta_{f_k}A_{i'j'k'}=3A_{k[j'k'}\delta_{i']}^{k+1},\n
&&\delta_{f_k}\varphi^{i'j'k'}=-3\varphi^{k+1[j'k'}\delta^{i']}_{k},\\
&&\delta_{f_k}V_{i'}^{a'}=\delta_{i'}^{k+1}\delta_{k}^{l'}V_{l'}^{a'},\n
&&\n
&&\delta_{h_k}A_{i'j'k'}=3(A_{k[j'k'}\delta_{i']}^k
                         -A_{k+1[j'k'}\delta_{i']}^{k+1}),\n
&&\delta_{h_k}\varphi^{i'j'k'}=-3(\varphi^{k[j'k'}\delta^{i']}_{k}
                                  -\varphi^{k+1[j'k'}\delta^{i']}_{k+1}),\\
&&\delta_{h_k}V_{i'}^{a'}=(\delta_{i'}^{k}\delta_{k}^{l'}
                           -\delta_{i'}^{k+1}\delta_{k+1}^{l'})V_{l'}^{a'}
\nonumber
\eeqa
$(k=1,\ldots,7)$, and 
\beqa
&&\delta_{e_8}A_{i'j'k'}=6\sqrt{3}\delta^{678}_{[i'j'k']},
\n
&&\delta_{e_8}\varphi^{i'j'k'}=\frac{1}{12}
\epsilon^{i'j'k'678p'q'r'}A_{p'q'r'},
\label{e8}\\
&&\delta_{e_8}V_{i'}^{a'}=
\left(-\frac{\sqrt{3}}{2}\varphi^{l'[78}\delta_{i'}^{6]}
+\frac{1}{432}\epsilon^{l'p'q'678s't'u'}A_{i'p'q'}A_{s't'u'}
\right)V_{l'}^{a'},\n
&&\n
&&\delta_{f_8}A_{i'j'k'}=-\frac{1}{6}
\epsilon_{i'j'k'678p'q'r'}\varphi^{p'q'r'}
%\n&&~~~~~~~~~~~~~~~~
-\frac{3\sqrt{3}}{2}\left(A_{i'[78}A_{6]j'k'}
-\frac{1}{9}A_{i'j'k'}A_{678}\right),\n
&&\delta_{f_8}\varphi^{i'j'k'}=-6\sqrt{3}\delta_{678}^{[i'j'k']}
+\frac{3\sqrt{3}}{2}\left(
A_{p'[78}\varphi^{p'[j'k'}\delta_{6]}^{i']}
-\frac{1}{9}A_{678}\varphi^{i'j'k'}
\right)\n&&~~~~~~~~~~~~~~~~
+\frac{1}{144}\epsilon^{i'j'k'p'q'r's't'u'}
A_{p'[78}A_{6]q'r'}A_{s't'u'},\label{f8}\\
&&\delta_{f_8}V_{i'}^{a'}=-\sqrt{3}\left(
\frac{1}{2}A_{i'[78}\delta_{6]}^{l'}
-\frac{1}{9}\delta_{i'}^{l'}A_{678}
\right)V_{l'}^{a'}\n
&&~~~~~~~~~~~~~~~
-\frac{\sqrt{3}}{18}\left(
A_{i'q'[6}A_{78]q'}+\frac{1}{4}A_{i'[78}A_{6]p'q'}
-\frac{1}{6}A_{i'p'q'}A_{678}
\right)\varphi^{l'p'q'}
,\n
&&\n
&&\delta_{h_8}A_{i'j'k'}= 
3A_{l'[j'k'}
\tilde{H}_{(8)i']}^{~~~~l'},\n
&&\delta_{h_8}\varphi^{i'j'k'}=
-3\varphi^{l'[j'k'}\tilde{H}_{(8)l'}^{~~~~i']},\label{h8}\\
&&\delta_{h_8}V_{i'}^{a'}=
\tilde{H}_{(8)i'}^{~~~~l'}V_{l'}^{a'}.\nonumber
\eeqa

Since $\delta_{f_8}$ makes $\varphi^{i'j'k'}$ deviate from the 
condition $\varphi^{i'j'k'}=0$ ($i',j',k'\neq9$), and consequently 
breaks the traceless condition for $V_{i'}^{a'}$, one needs a 
compensating local $SO(16)$ transformation. Such a transformation is 
given by 
\beqa
&&\delta_{f_8}^{\mbox{\scriptsize (comp)}}
{\cal V}=
{\cal V}Y,\n
&&
Y=\mbox{ad}[0,w_{a'b'c'},w^{a'b'c'}],\n
&&
w_{a'b'c'}=6\sqrt{3}\delta_{a'a}\delta_{b'b}\delta_{c'c}
e_6^ae_7^be_8^c.
\eeqa
The relation 
\beq
{\cal V}_-^{-1}
\delta_{f_8}^{\mbox{\scriptsize (comp)}}
{\cal V}_-+
\delta_{f_8}^{\mbox{\scriptsize (comp)}}
{\cal V}_+{\cal V}_+^{-1}
={\cal V}_+Y{\cal V}_+^{-1}
\eeq
enables us to find 
\beqa
&&\delta_{f_8}^{\mbox{\scriptsize (comp)}}
A_{i'j'k'}=6\sqrt{3}g_{i6}g_{j7}g_{k8}
\delta_{[i'}^i\delta_{j'}^j\delta_{k']'}^k,\n
&&\delta_{f_8}^{\mbox{\scriptsize (comp)}}
\varphi^{i'j'k'}=6\sqrt{3}
(\delta_6^{[i'}\delta_7^{j'}\delta_8^{k']'}
+3\psi_{[6}\delta_7^{[i'}\delta_{8]}^{j'}\delta_9^{k']'})
-\frac{1}{12}\epsilon^{i'j'k'pqrs't'u'}g_{p6}g_{q7}g_{r8}A_{a't'u'},
\n
&&\delta_{f_8}^{\mbox{\scriptsize (comp)}}
V_{i'}^{a'}=\left[
\frac{\sqrt{3}}{2}g_{i6}g_{j7}g_{k8}\delta_{i'}^{[i}\varphi^{jk]l'}
-\frac{\sqrt{3}}{2}(\delta_{[6}^{l'}A_{78]i'}
+\psi_{[6}A_{78]i'}\delta_9^{l'})
\right.\n
&&\left.~~~~~~~~~~~~~~~~~~+\frac{1}{432}\epsilon^{l'm'n'pqrs't'u'}
A_{i'm'n'}A_{s't'u'}g_{p6}g_{q7}g_{r8}
\right]V_{l'}^{a'}.
\eeqa
Thus the total transformation 
$\delta_{f_8}^{\mbox{\scriptsize (tot)}}\equiv
\delta_{f_8}+\delta_{f_8}^{\mbox{\scriptsize (comp)}}$
reads (taking the factor $-2\sqrt{3}$ in eqs.(\ref{normalization}) 
into account)
\beqa
\delta_{f_8}^{\mbox{\scriptsize (tot)}}A_{ijk}
&=&-3(g_{i6}g_{j7}g_{k8})_{[678]}
-\frac{1}{2}\epsilon_{ijk678pq}\varphi^{pq}\n
&&+9(A_{i78}A_{6jk})_{[ijk][678]}-A_{ijk}A_{678},\n
\delta_{f_8}^{\mbox{\scriptsize (tot)}}\varphi^{ij}
&=&-3\psi_{[6}\delta_7^i\delta_{8]}^j 
-\frac{1}{12}\epsilon^{ijpqrstu}g_{p6}g_{q7}g_{r8}A_{stu}\n
&&-(6A_{p[78}\varphi^{p[j}\delta_{6]}^{i]}-A_{678}\varphi^{ij})
+\frac{1}{12}\epsilon^{ijpqrstu}A_{p[78}A_{6]qr}A_{stu},\n
\delta_{f_8}^{\mbox{\scriptsize (tot)}}e_i^{~a}
&=&6(A_{i[78}e_{6]}^{~a}-\frac{1}{9}A_{678}e_i^{~a}),\n
\delta_{f_8}^{\mbox{\scriptsize (tot)}}\psi_i
&=&3(g_{i6}g_{j7}g_{k8})_{[678]}\varphi^{jk}+3A_{i[78}\psi_{6]}
-\frac{1}{36}\epsilon^{jkpqrstu}A_{ijk}A_{stu}g_{p6}g_{q7}g_{r8}\n
&&-(4A_{ij[6}A_{78]k}+A_{i[78}A_{6]jk}-\frac{2}{3}A_{ijk}A_{678})
\varphi^{jk}.
\label{f8tot}
\eeqa

We also write down $\delta_{e_8}$ on these fields:
\beqa
\delta_{e_8}A_{ijk}&=&-3\delta_{[i}^6\delta_j^7\delta_{k]}^8,\n
\delta_{e_8}\varphi^{ij}&=&
\frac{1}{12}\epsilon^{ij678pqr}A_{pqr},\n
\delta_{e_8}e_i^{~a}&=&0,\n
\delta_{e_8}\psi_i&=&-3\varphi^{[78}\delta_i^{6]}
-\frac{1}{36}\epsilon^{pq678stu}A_{ipq}A_{stu}.
\eeqa 

As a final comment we note that $\delta_{f_k}$, $k=1,\ldots,7$ 
break the upper-triangular gauge condition for $V_{i'}^{a'}$. 
In checking the $E_9$ relations in the next section this is 
harmless since $\delta_{f_k}V_{i'}^{a'}$ is well-defined whether 
or not $V_{i'}^{a'}$ is in upper-triangular form. In contrast, 
we have to compensate nonzero $\varphi^{678}$ as we did above 
because by definition one of $i',j',k'$ must be $9$ for 
$\varphi^{i'j'k'}$ and cannot be expressed in terms of 
$\varphi^{ij}$.

\section{$E_9$ in two dimensions}
\setcounter{equation}{0}
\subsection{``Matzner-Misner's'' $SL(2,\RRsubsection)$}
In this subsection we will analyze the extra $SL(2,\RR)$ 
symmetry that is newly present in two dimensions \footnote{The name 
``Matzner-Misner'' appeared in ref.\cite{Julia81}, in which the first 
evidence for $A_1^{(1)}$ including central charge was presented.}. 
As it mixes third 
and fourth rows of the elfbein, it is not manifest in the dualized 
Lagrangian (\ref{reducedL}) in which the coordinates $y$ and $x^1$ 
are differently treated. Hence we start with Cremmer-Julia's 
four-dimensional Lagrangian \cite{CJ} and then reduce the dimension 
to two.

The bosonic part of the four-dimensional Lagrangian obtained 
by dimensional reduction of the eleven-dimensional supergravity 
can be written in the form \cite{CJ}
\beq
{\cal L}'=E^{(4)}\left[
R^{(4)}
-\frac{1}{2}
 (\mbox{\boldmath $1$}^{MNPQ}m_{IJ}+\mbox{\boldmath $j$}^{MNPQ}a_{IJ})
 F_{MN}^I F_{PQ}^J
+\frac{1}{48}\mbox{Tr}G^{(4)MN}\partial_M{\cal R}^{-1}\partial_N{\cal R}
\right] 
\eeq
with
\beq
\mbox{\boldmath $1$}^{MNPQ}\equiv 
        G^{(4)P[M}G^{(4)N]Q},~
\mbox{\boldmath $j$}^{MNPQ}\equiv 
        \frac{1}{2}E^{(4)-1}\epsilon^{MNPQ}.
\eeq
$m_{IJ}$, $a_{IJ}$ are some symmetric functions of scalar fields. 
$\cal R$ is a symmetric matrix given in terms of $m$ and $a$ as  
\beq
\cal R=\left[
\begin{array}{cc}
m+am^{-1}a&am^{-1}\\m^{-1}a&m^{-1}
\end{array}
\right].
\eeq
$I,J$ are the internal indices labeling 28 
abelian vector fields. 
%%%%%%%%%%%%%%%%%%
\def\ov{\overline}
%%%%%%%%%%%%%%%%%%
$M,N,\ldots$ are the four-dimensional spacetime
indices, which we split into $(\ov{\mu},\ov{m})$
with $\ov{\mu}=t,x$ and $\ov{m}=y,x^1$. Assuming $\partial_{\ov{m}}=0$, 
we get the two-dimensional Lagrangian 
\beqa
{\cal L}'&=&\ov{e}\tilde{E}^{(2)}
\left[\tilde{R}^{(2)}+\tilde{G}^{(2)\mb\nb}
    \left(\frac{1}{4}\partial_{\mb}\ov{g}_{\ov{m}\ov{n}}
                     \partial_{\nb}\ov{g}^{\ov{m}\ov{n}}
          +\partial_{\mb}\ln\ov{e}\partial_{\nb}\ln\ov{e}
    \right)
\right.\n
&&~~~~~~~~~
-\frac{1}{4}\tilde{G}^{(2)\ov{\mu}\ov{\rho}}
            \tilde{G}^{(2)\ov{\nu}\ov{\sigma}}
\ov{g}_{\ov{m}\ov{n}}
\ov{B}^{\ov{m}}_{\mb\nb}\ov{B}^{\ov{n}}_{\ov{\rho}\ov{\sigma}}\n
&&~~~~~~~~~
-\frac{1}{2}m_{IJ}\left(
            \tilde{G}^{(2)\ov{\mu}\ov{\rho}}
            \tilde{G}^{(2)\ov{\nu}\ov{\sigma}}
            (F^I_{\mb\nb}-4F^I_{\mb\ov{m}}B_{\nb}^{\ov{m}})
             F^J_{\ov{\rho}\ov{\sigma}}
\right.\n
&&~~~~~~~~~
\left.
+2(\tilde{G}^{(2)\mb\nb}\ov{g}^{\ov{m}\ov{n}}
   +2\tilde{G}^{(2)\ov{\mu}[\ov{\nu}}\tilde{G}^{(2)\ov{\sigma}]\ov{\rho}}
    B_{\ov{\rho}}^{\ov{m}}B_{\ov{\sigma}}^{\ov{n}})
F^I_{\mb\ov{m}}F^J_{\nb\ov{n}}
\right)\n
&&~~~~~~~~~
+\ov{e}^{-1}(\tilde{E}^{(2)})^{-1}
a_{IJ}\epsilon^{\mb\nb}\epsilon^{\ov{m}\ov{n}}
F^I_{\mb\ov{m}}F^J_{\nb\ov{n}}\n
&&~~~~~~~~~
\left.
+\frac{1}{48}\mbox{Tr}\tilde{G}^{(2)\mb\nb}
\partial_{\mb}{\cal R}^{-1}\partial_{\nb}{\cal R}
\right],\label{anotherL_D=2}
\eeqa
where the vierbein is decomposed as
\beq
E_{~M}^{(4)A}
=\left[
\begin{array}{cc}
\tilde{E}_{~\mb}^{(2)\ab}
&\ov{B}_{\mb}^{\ov{m}}\ov{e}_{\ov{m}}^{\ab}\\
0&\ov{e}_{\ov{m}}^{\ab}
\end{array}
\right].
\eeq
The zweibein without tilde is reserved for the symbol used in the next 
section.
$E_{~M}^{(4)A}$ is related to the four-by-four block of the elfbein 
by 
\beq
E_{~M}^{(11)A}=
\left(\frac{e}{e_{i=1}^{a=1}}\right)^{-\frac{1}{2}}
E_{~M}^{(4)A}.
\eeq

The Lagrangian ${\cal L}'$ is manifestly invariant under 
the $SL(2,\RR)$ group transformation 
\beqa
\ov{e}_{\ov{m}}^{~\ov{a}}
        &\longmapsto&
               \Omega_{\ov{m}}^{~\ov{n}}\ov{e}_{\ov{n}}^{~\ov{a}},\n
F_{\mb\ov{m}}^{I}
        &\longmapsto&
               \Omega_{\ov{m}}^{~\ov{n}}F_{\mb\ov{n}}^{I},
\label{MMSL2R}
\\
\ov{B}_{\mb}^{~\ov{m}}
        &\longmapsto&
               (\Omega^{-1})_{\ov{n}}^{~\ov{m}}\ov{B}_{\mb}^{~\ov{n}}
\nonumber
\eeqa
for $\Omega_{\ov{m}}^{~\ov{n}}\in SL(2,\RR)$ group.
This is a generalization of Matzner-Misner's $SL(2,\RR)$ symmetry 
\cite{MM,Julia81,BM} in general relativity.

\subsection{From $E_8$ to $E_9$}
The Lagrangian ${\cal L}'$ differs from ${\cal L}$ 
by some other Lagrange multiplier terms, 
but their equations of motion are equivalent 
if the duality relations are used \cite{CJ}. Hence the equations 
of motion of ${\cal L}'$ are also equivalent to those derived 
from ${\cal L}+{\cal L}_{\mbox{\scriptsize Lag.mul.}}$ through 
the relations (\ref{duality1}) and (\ref{duality2}).
Thus the equations of motion of the eleven-dimensional supergravity 
possess $SL(2,\RR) \times$E${}_8$(group) symmetry if they are reduced to 
two dimensions. We will show that the infinitesimal form of this 
$SL(2,\RR)$ enlarges the symmetry algebra from $E_8$ to $E_9$.

Let $SL(2,\RR)_0$ be the infinitesimal form of (\ref{MMSL2R}), 
generated by
\beqa
\delta_X\ov{e}_{\ov{m}}^{~\ov{a}}
        &=&
               X_{\ov{m}}^{~\ov{n}}\ov{e}_{\ov{n}}^{~\ov{a}},\n
\delta_X F_{\mb\ov{m}}^{I}
        &=&
               X_{\ov{m}}^{~\ov{n}}F_{\mb\ov{n}}^{I},
\label{sl2R0}
\\
\delta_X\ov{B}_{\mb}^{~\ov{m}}
        &=&
               -X_{\ov{n}}^{~\ov{m}}\ov{B}_{\mb}^{~\ov{n}}.
\nonumber
\eeqa
We denote the generators for 
\beq
X=\left[\begin{array}{cc}&1\\0&\end{array}\right](\equiv E),~
\left[\begin{array}{cc}&0\\1&\end{array}\right](\equiv F)~\mbox{and}~
\left[\begin{array}{cc}1&\\&-1\end{array}\right](\equiv H)
\label{EFH}
\eeq
by $\delta_{e_0}$, $\delta_{f_0}$ and $\delta_{h_0}$, respectively.

We first note that the action of $SL(2,\RR)_0$ is expressed in terms 
of eleven-dimensional fields as 
\beqa
\delta_XE_{~~~\hat{\mu}}^{(11)\hat{\alpha}}
&=&\left[
\begin{array}{cccccc}
0&&&&&\\
&0&&&&\\
&&
\mbox{\Large $X$}\raisebox{-3mm}{\rule{0mm}{10mm}}
&&&\\
&&&0&&\\
&&&&\ddots&\\
&&&&&0
\end{array}
\right]\left[\begin{array}{c}
\raisebox{2mm}{$\vdots$}\\
E_{~~~y}^{(11)\hat{\alpha}}\\
E_{~~~1}^{(11)\hat{\alpha}}\\
\\
\vdots\\
\\
\end{array}\right]\n
&\equiv&X_{(0)\hat{\mu}}^{~~~~\hat{\nu}}E_{~~~\hat{\nu}}^{(11)\hat{\alpha}}
\label{XE11}
\eeqa
and 
\beq
\delta_XA_{\hat{\mu}\hat{\nu}\hat{\rho}}
=3(X_{(0)\hat{\mu}}^{~~~~\hat{\sigma}}
A_{\hat{\sigma}\hat{\nu}\hat{\rho}})_{[\hat{\mu}\hat{\nu}\hat{\rho}]}.
\eeq
In fact, if we define 
\beq
X_{(k)\hat{\mu}}^{~~~~\hat{\nu}}
=
\begin{array}{c}
{
    \begin{array}{ccccccccc}
    \mbox{~~\scriptsize $t$}
    &\mbox{\scriptsize $x$}
    &\mbox{\scriptsize $y$}
    &\mbox{\scriptsize $1$}
    &~\cdots
    &\mbox{\scriptsize $k$}
    &\!\!\mbox{\scriptsize $(\!k\!+\!1\!)$}
    &\!\cdots
    &\mbox{\scriptsize $8$}
\end{array}
}\\
{
    \left[\begin{array}{ccccccccc}
        0&&&&&&&&\\
        &0&&&&&&&\\
        &&0&&&&&&\\
        &&&0&&&&&\\
        &&&&\ddots&&&&\\
        &&&&&\mbox{~\Large $X$}&&&\\
        &&&&&&&&\\        
        &&&&&&&\ddots&\\
        &&&&&&&&0\\
    \end{array}\right]    
}
\end{array},
\eeq
the action of $SL(2,\RR)_k$ $(k=1,\ldots,7)$ can also 
be written as 
\beqa
\delta_{x_k}E_{~~~\hat{\mu}}^{(11)\hat{\alpha}}
&=&
X_{(k)\hat{\mu}}^{~~~~\hat{\nu}}E_{~~~\hat{\nu}}^{(11)\hat{\alpha}},
\label{deltaxkE}
\\
\delta_{x_k}A_{\hat{\mu}\hat{\nu}\hat{\rho}}
&=&3(X_{(k)\hat{\mu}}^{~~~~\hat{\sigma}}
A_{\hat{\sigma}\hat{\nu}\hat{\rho}})_{[\hat{\mu}\hat{\nu}\hat{\rho}]},
\eeqa
where $x=e,f,h$
corresponding to $X=E,F,H$. 
Hence $SL(2,\RR)_0$ commutes with $SL(2,\RR)_k$ $(k=2,\ldots,7)$. 
It is also easy to see that $SL(2,\RR)_0$ and $SL(2,\RR)_1$ do not 
commute but generate $SL(3,\RR)$. As we already remarked, 
(\ref{deltaxkE}) is well-defined irrespective of the decomposition 
of $E^{(11)\hat{\alpha}}_{~~~\hat{\mu}}$ or whether or not 
$E^{(3)\alpha}_{~~\mu}$ or $e_i^{~a}$ is in upper-triangular form. 
Thus the matter is reduced to a trival algebra of matrices. 
If the set $\{-\delta_{e_0},-\delta_{f_0},-\delta_{h_0}\}$
is regarded as a part of the Chevalley generators, 
the corresponding elements of the ``extended Cartan matrix'' are
\beq
K_{0k}=K_{k0}=2\delta_{k,0}-\delta_{k,1}~~~(k=0,\ldots,7).\label{K0k}
\eeq

Next consider the relation between $SL(2,\RR)_0$ and $SL(2,\RR)_8$. 
$\delta_{h_8}$-charges of the scalar fields in three dimensions 
can be read off from (\ref{h8}). Let $q(i)$ be a function defined by 
\beq
q(i)=\left\{
\begin{array}{rl}
\mbox{\Large $-\frac{1}{3}$}&i=1,\ldots,5,\\
&\\
\mbox{\Large $\frac{2}{3}$}&i=6,7,8,
\end{array}
\right.
\eeq
then $\delta_{h_8}$-charges $q_8$ are 
\beqa
q_8(e_i^{~a})&=&q(i),\n
q_8(\psi_{i})&=&\frac{1}{3}+q(i),\n
q_8(A_{ijk})&=&q(i)+q(j)+q(k),\\
q_8(\varphi^{ij})&=&\frac{1}{3}-q(i)-q(j).\nonumber
\eeqa
$E^{(3)\alpha}_{~\mu}$ is of course $\delta_{h_8}$-neutral.
Using the duality relation, $\delta_{h_8}$-charges of 
$E_{~~~y}^{(11)\alpha}$ and $A_{yjk}$ turn out to be
\beq
q_8(E_{~~~y}^{(11)\alpha})=-\frac{1}{3}
\eeq
and
\beq
q_8(A_{yjk})=-\frac{1}{3}+q(i)+q(j),
\eeq
which are equal to $q(e_{1}^{~\alpha})$ and 
$q(A_{1jk})$, respectively. 
Thus $SL(2,\RR)_0$ does not change $\delta_{h_8}$-charge.

Similarly, $\delta_{h_0}$-charges $q_0$ of the eleven-dimensional 
fields are 
\beq
q_0(E_{~~~\hat{\mu}}^{(11)\hat{\alpha}})=
+\delta_{\hat{\mu}}^{y}-\delta_{\hat{\mu}}^{1},~
q_0(A_{\hat{\mu}\hat{\nu}\hat{\rho}})=
(\delta_{\hat{\mu}}^{y}+\delta_{\hat{\nu}}^{y}+\delta_{\hat{\rho}}^{y})
-(\delta_{\hat{\mu}}^{1}+\delta_{\hat{\nu}}^{1}+\delta_{\hat{\rho}}^{1}).
\eeq
The duality relations translate these equations into those in terms   
the three- dimensional fields as follows:
\beqa
q_0(e_i^{~a})&=&-\delta_i^1,\n
q_0(\psi_{i})&=&-1-\delta_i^1,\n
q_0(A_{ijk})&=&-(\delta_i^1+\delta_j^1+\delta_k^1),\\
q_0(\varphi^{ij})&=&-1+\delta_i^1+\delta_j^1,\n
q_0(E_{~~\mu}^{(3)\alpha})&=&-1+\delta_{\mu}^y.\nonumber
\eeqa
In view of (\ref{e8})(\ref{f8}), we also find that $SL(2,\RR)_8$ 
does not change $\delta_{h_0}$-charge. 

\subsection{Serre relation}
What remains to check is the Serre relation between 
$SL(2,\RR)_0$ and $SL(2,\RR)_8$. 
Since $\delta_{f_0}$ gives rise to a non-zero $E^{(11)\alpha}_{~~~i}$ 
element, one needs a local Lorentz transformation 
$\delta_{f_0}^{\mbox{\scriptsize (comp)}}$ to restore the 
``block-upper-triangular'' form of the elfbein (\ref{3+8}). 
Including this contribution, the total variation  
$\delta_{f_0}^{\mbox{\scriptsize (tot)}}
\equiv\delta_{f_0}+\delta_{f_0}^{\mbox{\scriptsize (comp)}}$ is 
given as 
\beqa
\delta_{f_0}^{\mbox{\scriptsize (tot)}}
E^{(3)\alpha}_{~~\mu}&=&2B_{[y}^{~~1}E^{(3)\alpha}_{~~\mu]},\n
\delta_{f_0}^{\mbox{\scriptsize (tot)}}
B_{\mu}^{~i}&=&e^{-2}G^{(3)}_{\mu y}g^{i1}-B_{\mu}^{~1}B_{y}^{~i},\n
\delta_{f_0}^{\mbox{\scriptsize (tot)}}
e_i^{~a}&=&\delta_i^1B_{\mu}^{~j}e_j^{~a},\n
\delta_{f_0}^{\mbox{\scriptsize (tot)}}
A_{ijk}&=&3\delta_{[i}^1A_{jk]y},\n
\delta_{f_0}^{\mbox{\scriptsize (tot)}}
A_{\mu ij}&=&2\delta_{[i}^1A_{j]\mu y},\n
\delta_{f_0}^{\mbox{\scriptsize (tot)}}
A_{\mu\nu i}&=&\delta_{i}^1A_{\mu\nu y},\n
\delta_{f_0}^{\mbox{\scriptsize (tot)}}
A_{\mu\nu\rho}&=&0.
\eeqa
The rule of $\delta_{e_0}$ is much simpler:
\beqa
&
\delta_{e_0}E^{(3)\alpha}_{~~\mu}
=\delta_{e_0}e_i^{~a}
=\delta_{e_0}A_{ijk}=0,&\n
&
\delta_{e_0}B_{\mu}^{~i}
=\delta_\mu^y\delta_1^i,&\n
&
\delta_{e_0}A_{\mu ij}
=\delta_\mu^yA_{1ij},&\n
&
\delta_{e_0}A_{\mu\nu i}
=2\delta_{[\mu}^yA_{\nu]i1},&\n
&
\delta_{e_0}A_{\mu\nu\rho}
=3\delta_{[\mu}^yA_{\nu\rho]1}.&
\eeqa

To prove that the symmetry algebra is $E_9$ one needs to show 
$
[\delta_{e_0},~\delta_{e_8}]=
[\delta_{f_0},~\delta_{f_8}]=0
$ 
on all the fields that appear in the theory. 
What makes the situation difficult is the fact that 
$SL(2,\RR)_8$ is originally defined as a variation on the  
three-dimensional dualized fields $\varphi^{ij}$ and $\psi_i$ 
while $SL(2,\RR)_0$ naturally acts on the eleven-dimensional fields, 
or ``un-dualized fields'' $A'_{\mu ij}$ and $B_\mu^{~i}$.  
Thus all one can do is to verify the commutator to vanish only 
on their derivatives using the eqs. (\ref{duality1})(\ref{duality2}) 
\cite{NicolaiPL}. Since there appear only the field strengths 
in the original Lagrangian, the gauge potentials $A'_{\mu ij}$ and 
$B_\mu^{~i}$ themselves do not affect physics classically, but they 
can quantum mechanically. We will restrict ourselves to the 
classical aspect of $E_9$ in this paper.

Let us examine $[\delta_{f_0}^{\mbox{\scriptsize (tot)}}
,~\delta_{f_8}^{\mbox{\scriptsize (tot)}}]$. 
It is easy to find that 
$[\delta_{f_0}^{\mbox{\scriptsize (tot)}}
,~\delta_{f_8}^{\mbox{\scriptsize (tot)}}]e_i^{~a}
=[\delta_{f_0}^{\mbox{\scriptsize (tot)}}
,~\delta_{f_8}^{\mbox{\scriptsize (tot)}}]B_{\mu\nu}^i
=0$ if and only if 
\beqa
\delta_{f_8}^{\mbox{\scriptsize (tot)}}
B_y^{~i}&=&6A'_{y[78}\delta_{6]}^i,\n
\delta_{f_8}^{\mbox{\scriptsize (tot)}}
B_\mu^{~1}&=&0.
\eeqa 
They are consistent with the equation 
\beq
\delta_{f_8}^{\mbox{\scriptsize (tot)}}
B^k_{\mu\nu}=6F'_{\mu\nu[78}\delta_{6]}^k,
\eeq
which is derived from (\ref{f8tot}) and (\ref{duality2}).
Then $[\delta_{f_0}^{\mbox{\scriptsize (tot)}},
~\delta_{f_8}^{\mbox{\scriptsize (tot)}}]
E^{(3)\alpha}_{~~\mu}=0$ follows. 

Since $\delta_{f_0}^{\mbox{\scriptsize (tot)}}
\delta_{f_8}^{\mbox{\scriptsize (tot)}}A_{ijk}$ contains a term 
$-\frac{1}{2}\epsilon_{ijk678pq}\delta_{f_0}^{\mbox{\scriptsize (tot)}}
\varphi^{pq}$, we calculate the commutator on $\partial_{\mu}A_{ijk}$ 
as we noted above. Useful formulas are  
\beqa
\delta_{f_8}^{\mbox{\scriptsize (tot)}}
F_{\mu\nu}^{(3)ij}&=&-3B_{\mu\nu[6}\delta_7^{[i}\delta_{8]}^{j]}
-12(A_{p78}F_{\mu\nu}^{(3)p[j}\delta_6^{i]})_{[678]}
+\frac{2}{3}A_{678}F_{\mu\nu}^{(3)ij}\n
&&+\frac{1}{6}E^{(3)-1}e^{-2}\epsilon^{\mu\nu\rho}
\epsilon^{ijpqrstu}\partial_{\rho}A_{stu}g_{p6}g_{q7}g_{r8},\\
\delta_{f_0}^{\mbox{\scriptsize (tot)}}
(E^{(3)}e^2F^{(3)\bar{\mu}yij})
&=&E^{(3)}e^2\left[
2{F''}^{\bar{\mu}y~[j}_{~~~y}g^{i]1}
+B_y^{~1}F^{(3)\bar{\mu}yij}+(2B_y^{~j}F^{(3)\bar{\mu}y1i})^{[ij]}
\right.\n
&&\left.
-e^{-2}g^{ik}g^{jl}g^{1m}\partial^{\bar{\mu}}A_{klm}
\right],
\eeqa
where 
$F''_{~\mu\nu\rho i}\equiv
F_{\mu\nu\rho i}-3B_{[\mu}^{k}F_{\nu\rho]ki}
                +3B_{[\mu}^jB_{\nu}^kF_{\rho]jki}$ is the Kaluza-Klein 
invariant non-dynamical three-form field strength, which was ignored 
in the three-dimensional reduced Lagrangian (\ref{3dLag}).
Using these equations, one finds
\beq
[\delta_{f_0}^{\mbox{\scriptsize (tot)}}
,~\delta_{f_8}^{\mbox{\scriptsize (tot)}}]\partial_{\rho}A_{ijk}
=-\frac{3}{2}\delta_{[i}^1\epsilon_{jk]1678pq}\epsilon_{\mu\nu\rho}
E^{(3)}e^2 {F''}^{\mu\nu~[q}_{~~y}g^{p]1}.
\eeq
Thus the commutator can be consistently set to zero.

The commutator on $F^{(3)}_{\mu\nu ij}$ is the 
most complicated but can be done straightforwardly. In practice 
much labor can be saved if one calculates 
$[\delta_{f_0}^{\mbox{\scriptsize (tot)}}
,~\delta_{f_8}^{\mbox{\scriptsize (tot)}}](E^{(3)}e^2
F^{(3)\bar{\mu}yij})$ and uses the above formulas. It is worth 
noting that $F^{(3)\bar{\mu}\bar{\nu}ij}
=B^{\bar{\mu}\bar{\nu}}_k=0$ 
since $\partial_y=0$ for any field in rhs of the duality 
relations (\ref{duality1})(\ref{duality2}). In fact, this is a part 
of Geroch's compatibility condition considered in the next subsection.
After some calculation one may confirm that 
$[\delta_{f_0}^{\mbox{\scriptsize (tot)}}
,~\delta_{f_8}^{\mbox{\scriptsize (tot)}}](E^{(3)}e^2
F^{(3)\bar{\mu}yij})$ vanishes up to terms proportional to 
$F''_{\mu\nu\rho i}$.

Let us now turn to the examination of 
$[\delta_{e_0},~\delta_{e_8}]$. This vanishes on 
$E^{(3)\alpha}_{~~\mu}$, $e_i^{~a}$ and $A_{ijk}$ trivially. 
It is also easy to see that 
$\delta_{e_0}B^{\mu\nu}_k=\delta_{e_0}F^{(3)\mu\nu ij}=0$, 
so that 
\beqa
&&\delta_{e_0}\varphi^{ij}=c^{ij},\n
&&\delta_{e_0}(\partial_{\mu}\psi_i)=-c^{jk}A_{ijk}+d_i,
\eeqa
where $c^{ij}$, $d_i$ are constants. Thus $[\delta_{e_0},~\delta_{e_8}]$ 
vanishes also on $\partial_{\mu}\varphi^{ij}$ and 
$\partial_{\mu}\psi_i$. (One may even check that it vanishes 
on $\varphi^{ij}$, and does on $\psi_i$ provided that 
$c^{67}=c^{78}=c^{86}=0$.)

\begin{figure} 
\caption{Dynkin diagram of $E_9$.}
\phantom{space}
\centerline{
\mbox{\psfig{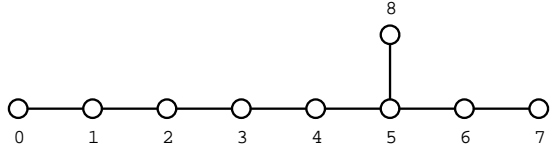}}
}
\label{E9Dynkin}
\end{figure}

Finally let us consider the non-dynamical fields (three- and four-form 
field strengths). One might worry that the $SL(2,\RR)_0$ may give rise 
to a dynamical field out of non-dynamical one, so that the latter may 
not be consistently ignored. Since 
$\delta_{f_0}^{\mbox{\scriptsize (tot)}}$ effectively replaces $i,j,\dots$ 
indices with $y$, only $\delta_{e_0}$ is dangerous. Happily, 
the Kaluza-Klein invariant combinations 
$A'_{\mu\nu i}\equiv 
A_{\mu\nu k}
-2B_{[\nu}^{~j}A_{\mu]jk}
+B_{[\mu}^{~i}B_{\nu]}^{~j}A_{ijk}$ 
and 
$A'_{\mu\nu\rho}\equiv A_{\mu\nu\rho}
-3B_{[\rho}^{~k}A_{\mu\nu]k}
+3B_{[\nu}^{~j}B_{\rho}^{~k}A_{\mu]jk}
-B_{[\mu}^{~i}B_{\nu}^{~j}B_{\rho]}^{~k}A_{ijk}$ 
are $\delta_{e_0}$-invariant, and so are 
the bilinear of 
$F''_{\mu\nu\rho k}$ ($=3\partial_{[\mu}A'_{\nu\rho]k}
-3B^j_{[\mu\nu}A'_{\rho]jk}$)
and 
$4\partial_{[\mu}A'_{\nu\rho\sigma]}
-6B^j_{[\mu\nu}A'_{\rho\sigma]k}$
omitted in the Lagrangian (\ref{3dLag}). 
Hence $\delta_{e_0}$ commutes with $\delta_{e_8}$ on these fields. 
This completes the proof of the commutativity of $SL(2,\RR)_0$ 
and $SL(2,\RR)_8$ and 
allows us to define 
\beq
K_{08}=K_{80}=0. \label{K08}
\eeq
(\ref{K0k}) and (\ref{K08}) extend the Cartan matrix (\ref{E8Cartan}) 
to
%
%\beq
%K_{ij}=\left[
%\begin{array}{c|cccccccc}
%2&-1&&&&&&&\\
%\hline 
%\!\!-1\!\!&2&\!\!-1\!\!&&&&&&\\
%&\!\!-1\!\!&2&\!\!-1\!\!&&&&&\\
%&&\!\!-1\!\!&2&\!\!-1\!\!&&&&\\
%&&&\!\!-1\!\!&2&\!\!-1\!\!&&&\\
%&&&&\!\!-1\!\!&2&\!\!-1\!\!&&\!\!-1\!\!\\
%&&&&&\!\!-1\!\!&2&\!\!-1\!\!&\\
%&&&&&&\!\!-1\!\!&2&\\
%&&&&&\!\!-1\!\!&&&2
%\end{array}
%\right],
%\eeq
%which is the Cartan matrix 
%
that 
of $E_9$ (Fig.\ref{E9Dynkin}).

\subsection{Compatibility of the symmetry 
with the Killing vector fields}
Let us conclude this section with a remark on Geroch's condition 
concerning the compatibility of the symmetry with the Killing vector 
fields \cite{Geroch2}. We have shown in this section that 
the new Killing vector field $\partial_y$ gives rise to an extra 
symmetry $SL(2,\RR)_0$ which together with the $E_8$ forms the 
$E_9$ commutation relations. For this $E_9$ to be really a symmetry 
of the reduced system, any Killing vector field assumed to exist 
must be mapped to another Killing vector field again by any action 
of the symmetry for consistency. In other word all the new fields 
after the transformation  must be again independent of $y$ 
or $x^i$, $i=1,\ldots, 8$. This is a trivial requirement if the 
infinitesimal transformation is given in terms of a function of 
eleven-dimensional fields only, but, if the variation includes dualized 
fields, it gives the constraints 
$\partial_y\psi_k=\partial_y\varphi^{ij}=0$, which imply 
\beq
B^{\ov{\mu}\ov{\nu}}_k=0,~~~F^{'\ov{\mu}\ov{\nu}ij}=0. 
\label{G-conditions1}
\eeq
Since the coordinates $y$ and $x^i$, $i=1,\ldots,8$ are now on the 
same footing, one may also take $y$ and seven of eight $x^i$ 
as the coordinates of an eight-torus, dualize the three-dimensional 
vectors and impose the independence of the remaining last one that 
was not taken as the coordinate of the eight-torus. One would then 
obtain similar conditions on the vector fields which may  
also have $y$ index. But the conditions (\ref{G-conditions1}) are 
based on a specific splitting of the three-dimensional space and 
do not allow a simple replacement of indices. To derive the fully 
$y$-$x^i$ symmetric conditions we rewrite (\ref{G-conditions1}) into 
equivalent expressions  
\beqa
&&\epsilon_{\hat{\mu}_0\hat{\mu}_1
\cdots\hat{\mu}_8
\hat{\nu}\hat{\rho}
}
\xi_{(0)}^{\hat{\mu}_0}
\xi_{(1)}^{\hat{\mu}_1}\cdots
\xi_{(8)}^{\hat{\mu}_8}
\nabla^{\hat{\nu}}
\xi_{(i)}^{\hat{\rho}}
=0,\n
&&\epsilon_{\hat{\mu}_0\hat{\mu}_1
\cdots\hat{\mu}_8
\hat{\nu}\hat{\rho}
}
\xi_{(0)}^{\hat{\mu}_0}
\xi_{(1)}^{\hat{\mu}_1}\cdots
\xi_{(8)}^{\hat{\mu}_8}
\xi_{(i)\hat{\tau}}
\xi_{(j)\hat{\sigma}}
\nabla^{[\hat{\nu}}
A^{\hat{\rho}\hat{\tau}\hat{\sigma}]}
=0, \label{G-conditions2}
\eeqa
where $\xi_{(i)}^{\hat{\mu}}=\delta_i^{\hat{\mu}}$ and 
$\xi_{(0)}^{\hat{\mu}}=\delta_y^{\hat{\mu}}$ are the Killing 
vector fields assumed to exist. They are manifestly $y$-$x^i$ 
symmetric if one allows $i$ to take $0$ and regards $x^0$ as $y$. 
With this extension (\ref{G-conditions2}) obviously generalize 
Geroch's compatibility condition derived in four-dimensional pure 
gravity \cite{Geroch2} and we assume them to hold for consistency. 
Note that the first set of equations are satisfied if all the Killing 
vector fields are hypersurface-orthogonal, while the second impose 
some conditions on the field strength of $A_{\mu ij}$ and 
$A_{\mu\nu i}$.

\section{$E_{10}$ in one dimension}
\setcounter{equation}{0}
\subsection{Null Killing vector in two dimensions}
So far we have discussed the $E_9$ symmetry in two dimensions.  
We would now like to study the enlargement of the symmetry 
to $E_{10}$. We first describe a special feature of dimensional 
reduction from two to one dimension \cite{NicolaiPL}.

We parameterize the dreibein as
\beq
E_{~\mu}^{(3)\alpha}=
\left[\begin{array}{cc}
E_{\mb}^{(2)\ab}&\rho A_{\mb}\\
0&\rho
\end{array}\right],
\eeq
where $\mu=\mb,y$ and $\mb=\dot{+}, \dot{-}$ 
(They are dotted in order to distinguish them from Lorentz indices.)
are three- and two- dimensional 
curved indices, whereas $\alpha=\ab,2$ and $\ab=+, -$ are corresponding  
Lorentz indices, respectively.  
(We switch the notation from $t,x$  to $\dot{+},\dot{-}$.)
Conventions for the flat metrics are 
\beq
\eta^{(3)}_{\alpha\beta}=
\left[\begin{array}{cc}
\eta^{(2)}_{\ab\beb}&\\
&1
\end{array}\right], ~~~
\eta^{(2)}_{\ab\beb}=
\left[\begin{array}{cc}
&1\\
1&
\end{array}\right].
\label{eta^3}
\eeq
Introducing the ninth Killing vector $\partial_y$, 
the three-dimensional nonlinear sigma model 
\beq
{\cal L}=E^{(3)}(R^{(3)}+\frac{1}{240}G^{\mu\nu}\mbox{Tr}
\partial_{\mu}{\cal M}^{-1}\partial_{\nu}{\cal M})
\eeq
is then reduced to 
\beq
{\cal L}=\rho E^{(2)}(R^{(2)}-\frac{1}{4}\rho^2H_{\mb\nb}H^{\mb\nb}
+\frac{1}{240}\mbox{Tr}\partial_{\mb}{\cal M}^{-1}\partial^{\mb}{\cal M}),
\label{L_D=2}
\eeq
where $H_{\mu\nu}=\partial_{\mb}A_{\nb}-\partial_{\nb}A_{\mb}$.
(\ref{L_D=2}) is on-shell equivalent to (\ref{anotherL_D=2}).

Next we take $\partial_{\mnd}$ as the tenth Killing vector. 
The duality relation (\ref{duality2}) then becomes
\beqa
B^i_{\mnd y}~=&0&=~E^{(3)}e^{-2}G^{(3)\pld\pld}
           g^{ij}(\partial_{\pld}\psi_j+\cdots),\\
B^i_{\pld \mnd}~=&\partial_{\pld}B^i_{\mnd}&=~E^{(3)}e^{-2}G^{(3)y\pld}
           g^{ij}(\partial_{\pld}\psi_j+\cdots),\\
B^i_{\pld y}~=&\partial_{\pld}B^i_y&=~-E^{(3)}e^{-2}G^{(3)\mnd\pld}
           g^{ij}(\partial_{\pld}\psi_j+\cdots).
\eeqa
The precise form of the terms represented by dots can be found in 
(\ref{duality2}). Suppose that $G^{(3)\pld\pld}\neq 0$. These equations 
require that $B^i_{\mnd}$ and $B^i_{y}$ are constants, and 
$\psi_i$, $\varphi^{ij}$ and $A_{ijk}$ are constrained by some relations 
before equations of motion are imposed. Then there is no degree of 
freedom for the dual field $\psi_i$ to carry. The duality relation 
ceases to relate the fields dual with each other and falls into 
trivial. A similar thing is true for the relation (\ref{duality1}). 
Therefore, one needs to take the Killing vector $\partial_{\mnd}$ to be 
{\em null-like} if one wants to have non-trivial duality relations
\cite{NicolaiPL}. 

We adopt the following parameterization for the zweibein 
\cite{BB,Dick,NicolaiNP}:
\beq
E^{(2)\ab}_{~\mb}=\left[\begin{array}{cc}
e_{\pld}^+&\mu_{\pld}^{\mnd}e_{\mnd}^-\\ 
\mu_{\mnd}^{\pld}e_{\pld}^+&e_{\mnd}^-
\end{array}\right].
\eeq
$\mu_{\pld}^{\mnd}$ and $\mu_{\mnd}^{\pld}$ are the Lorentzian analogue 
of Beltrami differentials. That is, for given 
$\mu_{\pld}^{\mnd}$ and $\mu_{\mnd}^{\pld}$, the line element 
\beq
ds^2=e_{\pld}^+e_{\mnd}^-
(dx^{\pld}+\mu_{\mnd}^{\pld}dx^{\mnd})
(dx^{\mnd}+\mu_{\pld}^{\mnd}dx^{\pld})
\eeq
can be cast into the form $2\lambda dudv$ for some conformal factor 
$\lambda$ and for some $u$, $v$ such that  
\beq
\mu_{\mnd}^{\pld}=\frac{\partial_{\mnd}u}{\partial_{\pld}u},~~~
\mu_{\pld}^{\mnd}=\frac{\partial_{\mnd}v}{\partial_{\pld}v}.
\eeq 
Hence $\mu_{\mnd}^{\pld}$ and $\mu_{\pld}^{\mnd}$ parameterize 
the ``conformal structure'' of two-dimensional Lorentzian metrics. 
Note that they are real and independent each other. 

The condition $G^{(3)\pld\pld}=0$ implies that 
$\mu_{\mnd}^{\pld}$ vanishes. 
The reduced Lagrangian reads 
\beqa
{\cal L}&=&E^{(2)-1}\left(-2\partial_{\pld}\rho
                \partial_{\pld}(\mu_{\mnd}^{\pld}e_{\pld}^+e_{\mnd}^-)
                +\frac{1}{2}\rho^3(\partial_{\pld}A_{\mnd})^2\right)\n
&&+\frac{1}{240}\rho
        \frac{2\mu_{\mnd}^{\pld}}{1-\mu_{\pld}^{\mnd}\mu_{\mnd}^{\pld}}
        \mbox{\normalsize Tr}
        ({\cal M}^{-1}\partial_{\pld}{\cal M})^2.
\eeqa
$\mu_{\mnd}^{\pld}$ is set to be zero after deriving  
the equations of motion. The independent ones are 
\beq
e_{\pld}^+e_{\mnd}^-
\partial_{\pld}\frac{\partial_{\pld}\rho}{e_{\pld}^+e_{\mnd}^-}
+\frac{1}{240}\rho
\mbox{Tr}({\cal M}^{-1}\partial_{\pld}{\cal M})^2=0\label{EM1}
\eeq
and
\beq
\partial_{\pld}A_{\mnd}=0.\label{EM2}
\eeq

\subsection{New $SL(2,\RRsubsection)$ symmetry}
We will now describe a new $SL(2,\RR)$ symmetry in one dimension.  
This $SL(2,\RR)$ is defined to act on the second and third 
rows of the elfbein. For $E_{~\mu}^{(3)\alpha}$ 
in the decomposition (\ref{3+8}) it only affects the lower-right
two-by-two part (before compensating gauge transformations). 
$e_i^a$ is left unchanged. Denoting 
the infinitesimal transformations corresponding to the 
Chevalley generators by 
$\{-\delta_{e_{-1}},-\delta_{f_{-1}},-\delta_{h_{-1}} \}$, 
the first and the last are given by 
\beq
\delta_{e_{-1}}\left[
\begin{array}{cc}e_{\mnd}^-&\rho A_{\mnd}\\0&\rho\end{array}
\right]
=
\left[
\begin{array}{cc}&1\\0&\end{array}
\right]
\left[
\begin{array}{cc}e_{\mnd}^-&\rho A_{\mnd}\\0&\rho\end{array}
\right]
=
\left[
\begin{array}{cc}0&\rho \\0&0\end{array}
\right],
\eeq
\beq
\delta_{h_{-1}}\left[
\begin{array}{cc}e_{\mnd}^-&\rho A_{\mnd}\\0&\rho\end{array}
\right]
=
\left[
\begin{array}{cc}1&\\&-1\end{array}
\right]
\left[
\begin{array}{cc}e_{\mnd}^-&\rho A_{\mnd}\\0&\rho\end{array}
\right]
=
\left[
\begin{array}{cc}e_{\mnd}^-&\rho A_{\mnd}\\0&-\rho\end{array}
\right],
\eeq
while the multiplication of $\left[
\begin{array}{cc}&0\\1&\end{array}
\right]$ does not preserve the upper-triangular gauge, 
so a suitable compensating gauge tranfomation is needed for 
$\delta_{f_{-1}}$ :
\beqa
\delta_{f_{-1}}
\left[\begin{array}{cc}
\phantom{\raisebox{-3mm}{$\rule{1mm}{8mm}$}}E_{\mb}^{(2)\ab}&\rho A_{\mb}\\
\begin{array}{cc}0&0\end{array}
&\rho
\end{array}\right]
&=&
\left[\begin{array}{cc}
 \begin{array}{cc}0&0\\0&0\end{array} &  \begin{array}{c}0\\0\end{array} \\
\begin{array}{cc}0&1\end{array} &  0 \\
\end{array}\right] 
\left[\begin{array}{cc}
\phantom{\raisebox{-3mm}{$\rule{1mm}{8mm}$}}E_{\mb}^{(2)\ab}&\rho A_{\mb}\\
\begin{array}{cc}0&0\end{array}
&\rho
\end{array}\right]
\n
&&+
\left[\begin{array}{cc}
\phantom{\raisebox{-3mm}{$\rule{1mm}{8mm}$}}E_{\mb}^{(2)\ab}&\rho A_{\mb}\\
\begin{array}{cc}0&0\end{array}
&\rho
\end{array}\right]\cdot \rho^{-1}e_{\mnd}^-
\left[\begin{array}{ccc}
0&0&1\\0&0&0\\0&-1&0\end{array}\right] \n
&=&
\left[
\begin{array}{ccc}
0&-e_{\mnd}^-A_{\pld}&\rho^{-1}e_{\pld}^+e_{\mnd}^-\\
0&-e_{\mnd}^-A_{\mnd}&0\\
0&0& \rho A_{\mnd}
\end{array}
\right].
\eeqa
These equations determine the transformation law of the components of 
the dreibein. We also define that the three-dimensional sigma-model 
scalars are invariant. These new $SL(2,\RR)$ transformations are 
summarized in Table {\ref{tableN}}. Note that $\mu_{\dot{-}}^{\dot{+}}$ 
is kept unchanged under these transformations, which is consistent 
with our assumption.
It is easy to see that the variations of the equations of motion 
(\ref{EM1})(\ref{EM2}) vanish if they are used themselves. Hence 
this $SL(2,\RR)$ is a symmetry.

%%%%%%%%%%%%%%%%%
\begin{table}
\caption{$SL(2,\RR)_{-1}$ transformations.}
\centerline{
\begin{tabular}{c|c|c|c|}
&$\delta_{e_{-1}}$&$\delta_{h_{-1}}$&$\delta_{f_{-1}}$\\
\hline
$\ep$&0&0&0\\
$\emi$&0&$\emi$&$-A_{\mnd}\emi$\\
$\mu_{\pld}^{\mnd}$&0&$-\mu_{\pld}^{\mnd}$
            &$-A_{\pld}+\mu_{\pld}^{\mnd}A_{\mnd}$\\
$\mu_{\mnd}^{\pld}$&0&0&0\\
$A_{\pld}$&0&$A_{\pld}$&$-A_{\pld}A_{\mnd}+\rho^{-1}e_{\pld}^+e_{\mnd}^-$\\
$A_{\mnd}$&1&$2A_{\mnd}$&$-A_{\mnd}^2$\\
$\rho$&0&$-\rho$&$\rho A_{\mnd}$\\
$e_i^a$&0&0&0\\
$\psi_i$&0&0&0\\
%$e_a^i$&0&0&0\\
%$B_{\pld}^i$&0&0&0\\
%$B_{\mnd}^i$&$B_y^i$&$B_{\mnd}^i$&0\\
%$B_y^i$&0&$-B_y^i$&$B_{\mnd}^i$\\
$A_{ijk}$&0&0&0\\
$\varphi^{ij}$&0&0&0\\
%$A_{\pld jk}$&0&0&0\\
%$A_{\mnd jk}$&$A_{yjk}$&$A_{\mnd jk}$&0\\
%$A_{yjk}$&0&$-A_{yjk}$&$A_{\mnd jk}$
\end{tabular}}
\label{tableN}
\end{table}
%%%%%%%%%%%%%%%%

For the fields ``before the dualization'' $B_{\mu}^i$ and $A_{\mu ij}$, 
the transformaions are defined as  
\beq
\delta_{e_{-1}}B_{\mnd}^i=B_y^i,~~
\delta_{h_{-1}}B_{\mnd}^i=B_{\mnd}^i,~~
\delta_{f_{-1}}B_{\mnd}^i=0,
\eeq
\beq
\delta_{e_{-1}}B_y^i=0,~~
\delta_{h_{-1}}B_y^i=-B_y^i,~~
\delta_{f_{-1}}B_y^i=B_{\mnd}^i,
\eeq
\beq
\delta_{e_{-1}}B_{\pld}^i=
\delta_{h_{-1}}B_{\pld}^i=
\delta_{f_{-1}}B_{\pld}^i=0,
\eeq
and 
\beq
\delta_{e_{-1}}A_{\mnd jk}=A_{yjk},~~
\delta_{h_{-1}}A_{\mnd jk}=A_{\mnd jk},~~
\delta_{f_{-1}}A_{\mnd jk}=0,
\eeq
\beq
\delta_{e_{-1}}A_{yjk}=0,~~
\delta_{h_{-1}}A_{yjk}=-A_{yjk},~~
\delta_{f_{-1}}A_{yjk}=A_{\mnd jk},
\eeq
\beq
\delta_{e_{-1}}A_{\pld jk}=
\delta_{h_{-1}}A_{\pld jk}=
\delta_{f_{-1}}A_{\pld jk}=0.
\eeq

The Killing vector $\partial_{\dot{-}}$ gives arise to additional 
Geroch compatibility conditions 
$B^{\dot{+}y}_k = F^{'\dot{+}yij}=0$. They are similarly expressed 
in the form (\ref{G-conditions2}) with one of $0,\ldots,8$ being 
replaced by $-1$, where 
$\xi_{(-1)}^{\hat{\mu}}\partial_{\hat{\mu}}=\partial_{\dot{-}}$.

\subsection{Non-trivial realization of the new $SL(2,\RRsubsection)$}
In the previous subsection we saw an extra $SL(2,\RR)$ symmetry  
in one dimension with keeping the degrees of freedom of Beltrami 
differentials. In this subsection we show that this transformation 
certainly enlarges the symmetry which already exists 
in higher dimensions. For this purpose we further compactify 
the $x^{\pld}$ direction on a circle and verify that the new $SL(2,\RR)$
includes transformations that change the conformal structure 
of the $(x^{\pld},x^{\mnd})$-torus. 
The transformation formula obtained in the last subsection shows that 
$\mu_{\pld}^{\mnd}$ does change under $\delta_{f_{-1}}$. We will 
show that this variation cannot be generated 
by a reparameterization. 

Before and after the action of $\delta_{f_{-1}}$, the Killing vector 
$\partial_{\mnd}$ is preserved. Such diffeomorphisms are caused 
by the vector fields of the form 
\beq
Y=\epsilon^{\pld}(x^{\pld})\partial_{\pld}+
\epsilon^{\mnd}(x^{\pld})\partial_{\mnd},
\eeq 
where $\epsilon^{\dot{\pm}}(x^{\pld})$ is periodic functions 
of $x^{\pld}$ only. 

A variation of a differential form under a diffeomorphism is given 
by its Lie derivative, that is 
\beqa
L_Y E_{\nb}^{(2)\ab}&=&\epsilon^{\mb}\partial_{\mb}E_{\nb}^{(2)\ab}
+E_{\mb}^{(2)\ab}\partial_{\nb}\epsilon^{\mb}\n
&=&\epsilon^{\pld}\partial_{\pld}E_{\nb}^{(2)\ab}
+E_{\mb}^{(2)\ab}\partial_{\pld}\epsilon^{\mb}\delta_{\nb}^{\pld},
\eeqa
where we used 
$\partial_{\mnd}E_{\nb}^{(2)\ab}=\partial_{\mnd}\epsilon_{\mb}=0$.
Since 
\beq
L_Y E_{\pld}^{(2)-}=\mu_{\pld}^{\mnd}L_Y E_{\mnd}^{(2)-}
+E_{\mnd}^{(2)-}L_Y\mu_{\pld}^{\mnd}, 
\eeq
the Lie derivative of $\mu_{\pld}^{\mnd}$ reads 
\beq
L_Y\mu_{\pld}^{\mnd}=\partial_{\pld}(\epsilon^{\pld}\mu_{\pld}^{\mnd}
+\epsilon^{\mnd}).
\eeq
Thus the change of $\mu_{\pld}^{\mnd}$ under the 
diffeomorphism is a total derivative, and in particular 
\beq
\int_{\mbox{\scriptsize period}}dx^{\pld}L_Y\mu_{\pld}^{\mnd}=0.
\eeq
On the other hand, one can obviously take $A_{\pld}$ and 
$\mu_{\pld}^{\mnd}$ such that they satisfy 
\footnote{
Being on a torus, the constant mode of $A_{\dot{+}}$ cannot 
be gauged away by the Kaluza-Klein gauge transformation.
}
\beq
\int_{\mbox{\scriptsize period}}dx^{\pld}(A_{\pld}-\mu_{\pld}^{\mnd}A_{\mnd})
\neq 0.
\eeq
Hence $\delta_{f_{-1}}\mu_{\pld}^{\mnd}$ cannot be generated by 
any diffeomorphism. This proves that the new $SL(2,\RR)$ is 
indeed an enlargement of the symmetry because all the rest do not 
affect the conformal structure of the $(x^{\pld},x^{\mnd})$-torus. 

\subsection{From $E_9$ to $E_{10}$} 
We will now focus on the question of the $E_{10}$ symmetry. 
Since $SL(2,\RR)_{-1}$ is so constructed 
that it acts on the eleven-dimensional fields as 
\beqa
\delta_{x_{-1}}E_{~~~\hat{\mu}}^{(11)\hat{\alpha}}
&=&
X_{(-1)\hat{\mu}}^{~~~~\hat{\nu}}E_{~~~\hat{\nu}}^{(11)\hat{\alpha}},
\\
\delta_{x_{-1}}A_{\hat{\mu}\hat{\nu}\hat{\rho}}
&=&3(X_{(-1)\hat{\mu}}^{~~~~\hat{\sigma}}
A_{\hat{\sigma}\hat{\nu}\hat{\rho}})_{[\hat{\mu}\hat{\nu}\hat{\rho}]}
\eeqa
($x=e,f,h$ and $X=E,F,H$ defined by (\ref{EFH})),
with 
\beq
X_{(-1)\hat{\mu}}^{~~~~~~\hat{\nu}}
=
\begin{array}{c}
{
    \begin{array}{cccccc}
    \mbox{\tiny $\dot{+}$}
    &\mbox{\tiny $\dot{-}$}
    &\mbox{\scriptsize $y$}
    &\mbox{\scriptsize $1$}
    &\mbox{$\cdots$}
    &\mbox{\scriptsize $8$}

\end{array}
}\\
{
    \left[\begin{array}{ccccc}
        0&&&&\\
        &\mbox{\Large $X$}&&&\\
        &&~\raisebox{-1mm}{0}&&\\
        &&&\ddots&\\
        &&&&0
    \end{array}\right]    
}
\end{array},
\eeq
it commutes with $SL(2,\RR)_{k}$ $(k=1,\ldots,7)$
and generates $SL(3,\RR)$ with $SL(2,\RR)_0$. It is also obvious 
that $SL(2,\RR)_{-1}$ and $SL(2,\RR)_8$ commute. Thus, regarding
$\{-\delta_{e_{-1}}$, $-\delta_{f_{-1}}$, $-\delta_{h_{-1}}\}$ as 
a new set of the Chevalley generators, one may 
extend the Cartan matrix as 
\beq
K_{-1~j}=K_{j~-1}=2\delta_{j,-1}-\delta_{j,0}
\eeq
$(j=-1,0,1,\ldots,8)$, giving the Cartan matrix of $E_{10}$.
This completes the proof of the main assertion of this paper.

Finally we would like to emphasize that the greatest difficulty 
in establishing the $E_{10}$ symmetry was (apart from the issue of 
non-trivial realization), after all, the check of the commutativity 
of $SL(2,\RR)_0$ and $SL(2,\RR)_8$ in the previous section, and once 
we proved $E_{9}$, it was rather easy to confirm $E_{10}$. This was 
because $SL(2,\RR)_{-1}$ manifestly commutes with the $E_8$ and hence 
one has only to examine the relation with $SL(2,\RR)_0$ to show that 
they generate $SL(3,\RR)$. Realized linearly, the latter was shown 
as a consequence of a trivial algebra of matrices.

\section{Conclusions}
\setcounter{equation}{0}

We have considered dimensional reduction of the eleven-dimensional 
supergravity to three, two and one dimension(s). We derived 
the three-dimensional $E_{8(+8)}/SO(16)$ nonlinear sigma model 
by direct dimensional reduction from eleven dimensions. Freudenthal's 
classical construction of $E_8$ turned out to reflect its 
``supergravity structure'' very clearly. 
In two dimensions we found a Matzner-Misner-type $SL(2,\RR)$ 
symmetry. The transformation rules corresponding to the Chevalley 
generators of $E_9$ were explicitly written down. We gave a complete 
check of the generating relations of $E_9$ on all the fields including 
the field strengths of $U(1)$ gauge fields, but the gauge potentials. 
This provides a proof of the $E_9$ (``$E_{9(+9)}$'') hidden symmetry 
of the bosonic part of the eleven-dimensional supergravity upon 
reduction to two dimensions. The generalized Geroch compatibility 
(hypersurface-orthogonality) condition was derived. The check on the 
gauge potentials is basically impossible in our scheme since the 
duality relations are only defined for their field strengths. Therefore  
our proof holds only classically. It is naturally expected that in the 
full quantum M theory only its discrete subgroup of $E_9$ should 
survive as an exact symmetry.   

Upon further reduction to one dimension we found a new 
$SL(2,\RRabstract)$ symmetry, the transformation rule of which is 
defined similarly to that of \cite{NicolaiPL} found in $D=4$ pure 
gravity. We had to take the Killing vector to be null so as for 
the duality relations to be non-trivial. Consequently, this 
$SL(2,\RRabstract)$ acts on the space of certain plane wave solutions 
propagating at the speed of light. The $E_9$ being established 
in two dimensions, it was not difficult to see that the full symmetry 
algebra is a real form of $E_{10}$. To show that this 
$SL(2,\RRabstract)$ cannot be expressed in terms of the old $E_9$ but 
truly enlarges the symmetry, we compactified the final two dimensions 
on a two-torus and confirmed that it changes the conformal structure 
of this two-torus.

There remain many things to be done for a better understanding 
of the $E_{10}$ symmetry. We conclude this paper by listing 
some of them: \\
\noindent
(i) It is a bit awkward to compactify the final two dimensions 
because this means that all the dimensions are compactified in 
typeII string theory. As already mentioned in the introduction, 
a symmetry truly larger than $E_9$ could be realized on the fields 
of three-form origin without compactifying the last dimension. 
\\
\noindent
(ii) One is tempted to think that $E_{10}$, which is a consequence 
of a null reduction, might be related to Matrix theory \cite{Matrix} 
proposed as a description of M theory in infinite momentum frame. 
A deeper understanding of null reduction \cite{JN,GM} is desirable. 
\\
\noindent
(iii) U duality of typeII string and M theory below three dimensions 
must be investigated.
\\
\noindent
(iv) Obviously, fermions must also be included.

\section*{Acknowledgments}
The author is grateful to H. Nicolai for many valuable discussions 
and comments, without which he could have never completed this work, 
and to R. W. Gebert and T. Kawai for helpful discussions. Special 
thanks go to E. Witten for discussions on the issue of non-trivial 
realization of $E_{10}$. He also wishes to thank the Alexander von 
Humboldt Foundation for financial support, and the Institute for 
Advanced Study in Princeton for hospitality and support.

\end{document}